\documentclass[physrev, twocolumn, amsmath, amssymb, floats, floatfix,longbibliography]{revtex4-2}
\usepackage{graphicx}
\usepackage[breaklinks,pdfborder={0 0 0},colorlinks=true,citecolor=blue,urlcolor=blue,linkcolor=blue]{hyperref}
\usepackage{braket}
\usepackage{bbm}
\usepackage{amstext}

\newcommand{\sign}{\mathrm{sgn}}
\newcommand{\Imag}{\mathrm{Im}}
\newcommand{\Real}{\mathrm{Re}}
\newcommand{\mean}[1]{\langle #1 \rangle}
\newcommand{\bmean}[1]{\bigl\langle #1 \bigr\rangle}
\newcommand{\Tr}{\mathrm{Tr}}

\begin{document}

\title[Spacetime propagators of massive, real scalar fields in 1D]{Exact results, transient generalized Gibbs ensembles, and analytic approximations
for spacetime propagators of massive, real scalar fields in one spatial dimension}

\author{Tobias Boorman}
\affiliation{SUPA, School of Physics and Astronomy, University of St.\ Andrews, St.\ Andrews KY16 9SS, United Kingdom}

\author{Bernd Braunecker}
\affiliation{SUPA, School of Physics and Astronomy, University of St.\ Andrews, St.\ Andrews KY16 9SS, United Kingdom}

\begin{abstract}
The massive, real scalar field described by the Klein-Gordon equation in one spatial
dimension is the most elementary example of a bosonic quantum field theory. It has been investigated for many
decades either as a simple academic theory or as a realistic emergent many-body theory in low-dimensional
systems. Despite this, the space and time behavior of its propagators have rarely been
in the foreground, and although exact results are known, there remain gaps in the description and a lack of an
in-depth physical analysis.
The aim of this paper is to address the deficits by providing a comprehensive discussion of the results,
and to show that this old theory still allows for several new results and insights.
To start, we review the known results by providing a rederivation in full detail, to which we add a
discussion on how exactly space and time variables
need to be extended to complex values to ensure analyticity throughout spacetime. This procedure shows
also how singularities on the lightcone need to be regularized to remain compatible with the analyticity and the physical
limit of a vanishing mass.
An extension to nonzero temperatures is provided by considering the contact of the field to a
nonrelativistic thermal reservoir, such as is necessary for emerging field theories in condensed matter systems.
Subsequently, it is shown that the transient, short spacetime propagation can be understood in the context
of the modern development of a generalized Gibbs ensemble, which describes a massless theory with an effective
temperature that is set by the Klein-Gordon mass and the physical temperature.
Finally, an approximation scheme is presented that captures the non-trivial
mass dependence of the propagators throughout all spacetime but involves only elementary functions.
\end{abstract}

\maketitle


\section{Introduction}

The massive, real scalar field described by the Klein-Gordon equation  \cite{Klein1926,Fock1926,Gordon1926} is the
most elementary quantum field theory. As such it has been widely studied across the literature, but if a further dimensional
restriction to one spatial dimension is introduced, then as an
elementary quantum field theory this becomes a rather academic example
that rarely receives primary attention.
This neglects however developments that provide a realistic physical
underpinning to such 1+1 dimensional theories (for one spatial and one temporal dimension).
On the one hand, it has been recently proposed \cite{Carlip2012} that the spacetime structure of the
universe may indeed be 1+1 dimensional near the Planck scale, so that any quantitative result
could be applied to test this proposal.
On the other hand, such a field
theory can also arise as an emergent, realistic low-energy description of a many-body system, and through appropriate
confinement the theory can indeed be entirely one-dimensional. Yet as emergent theories arise primarily near
quantum criticality the resulting one-dimensional fields are usually massless, and the opening of a mass
gap is of interest mainly to characterize a phase transition rather than providing a complete description of
the dynamics of the fluctuating fields after the transition. To characterize a gapped phase, it is for instance
sufficient to verify the exponential decay of correlation functions with increasing distance, whereas
the exact shape at short distances or in the time direction is of less relevance. In a discussion where such
results are just means to an end, there is no need to investigate such functions any further. We therefore perceive
a lack of proper attention given to the shape of propagators under the massive
one-dimensional Klein-Gordon equation as a function of all times $t$ and positions $x$, independent of their magnitudes.
The conceptual and technical difficulties arise primarily in the $(t,x)$ description, while the results
for frequency-momentum $(\omega,k)$ or time-momentum $(t,k)$ space are straightforward,
at least without coupling to a thermal reservoir.
In $(t,x)$ space, however, causality and Lorentz invariance impose rather subtle conditions on small
shifts of spacetime distances and on a singular behavior on the light cone that without the shifts
leaves room for ambiguity on the choice of the branch cut structure of the underlying analytic solutions.

To give an overview of the lack of complete discussion, we briefly summarize the corresponding
presentations in a number of textbooks. Most of the discussions apply to 3+1 dimensions, but the
underlying structure is similar to 1+1 dimensions. None of the references address temperature effects.
Peskin and Schroeder \cite{Peskin1995} provide an extensive discussion of the Klein-Gordon
propagators but keep them in integral form without evaluation and comment on the asymptotics. Similarly, in
Mandl and Shaw \cite{Mandl2010}, Nair \cite{Nair2005}, and Veltman \cite{Veltman1994} the integrals are stated but not evaluated.
Zinn-Justin \cite{Zinn-Justin2002} considers the $(t,k)$ and $(\omega,k)$ dependences but no $x$ dependence.
Weinberg \cite{Weinberg2005} discusses and writes down the integral forms for the $(t,x)$ dependence
but does not evaluate them fully.
Kaku \cite{Kaku1993} provides a solution in the form of the Bessel function $J_0$, and Huang \cite{Huang2010} and
Roman \cite{Roman1969} provide solutions in the forms of various Bessel functions. A further
discussion of their implications is not given though, and rather entertaining in this regard is also the comment
by Itzykson and Zuber \cite{Itzykson1980} after their Eq.~(1-170) that ``the explicit expressions involving Bessel
functions are not too illuminating''.
Greiner and Reinhard \cite{Greiner1992} instead derive the Bessel function expressions and discuss their asymptotics,
yet have a singular expression at the origin that according to Ref.~\cite{Zhang2010} should be absent.
Roepstorff \cite{Roepstorff1994} and Mussardo \cite{Mussardo2020} also derive results in terms of modified Bessel functions
and discuss briefly their asymptotic forms.

In research papers explicit results are mostly used as means to an end.
There are some focused excellent discussions though, such as by Ref.~\cite{diSessa1974} which provides a further
expansion of the explicit results into harmonic modes, or the more recent systematic rederivations by Ref.~\cite{Zhang2010}.
From the condensed matter background, Refs.~\cite{Starykh1999,Gulacsi1994,Voit1998} offer a thorough discussion
of gapped phases in quantum wires, parts of which we push further in the present paper.
On a technical level, Pais and Uhlenbeck \cite{Pais1950} provide an early reference to using hyperbolic coordinates to
compute the field theory's momentum integral.

Much is therefore known and, to cite Giambiagi \cite{Giambiagi1994}: ``It is very difficult to discuss the [\textellipsis\!]\
Klein-Gordon equation [\textellipsis\!]\ and say something new about it.''
This is the challenge we accept in this paper.
Even the dedicated literature either lacks to address comprehensively the
underlying analytic structure in the complex plane, does not provide closed solutions, or does not
exhibit the details of the derivation. As the analytic structure incorporates the causal behavior of
the theory as well as the qualitative change between a massive and a massless theory, we deemed a
thorough analysis necessary, filling the remaining gaps.
Furthermore we remark that none of the references addresses the effect
of a thermal bath on the massive system. This is of importance primarily for a condensed matter system,
but as we shall show, the transient behavior of the correlators even at zero temperature
can be understood in terms of an effective massless theory but at an effective temperature that is set by the mass.
The latter is an expression of a generalized Gibbs ensemble (GGE) \cite{Rigol2007,Caux2012,Vidmar2016},
and thus there exists a transient correspondence
between the massive Klein-Gordon theory and the modern development of GGEs.
To understand this correspondence we extend the theory to finite temperatures and investigate its properties
in various limits of temperature and mass.
All this together reveals a considerable amount of physical properties and subtleties
in the exact solutions that have been overlooked so far and whose discussion is, indeed, illuminating.

We first present the full solution of the propagators in $(t,x)$
space, building on the elementary Wightman functions but constructing with them the set of retarded, advanced, and
time ordered Green's functions. This reproduces known results, in the form of Bessel functions,
but complements them with the discussion of
analytic continuity across the light cone. We also show how the Lorentz invariance of the Klein-Gordon
equation explicitly appears in the propagators instead of just assuming that it has to be so.
We provide an extensive discussion of the physically most meaningful approach
to regularize the singularity on the light cone that appears in propagators with fixed
or time-ordered operator ordering but not in retarded and advanced functions. We argue that only the introduction
of a high energy cutoff is consistent with the analytic properties and physical limits, and that it preserves a
simpler mathematical form than other regularization schemes.

The result is subsequently generalized
to nonzero temperatures $T$ induced by the contact to a thermal reservoir. We show that an analytic
continuation can be applied such that this explicitly results in the necessary periodicity in the imaginary
time direction through a full summation over inverse Matsubara frequencies. This result provides in addition a
quantitative expression showing how exactly temperature corrections become exponentially small when the thermal
energy $k_B T$ (with $k_B$ the Boltzmann constant) is smaller than the energy set by the mass.
As an opposite situation we then provide a careful
investigation of the limit of vanishing mass. This limit is naturally singular because even an infinitesimal
mass qualitatively changes the underlying wave functions, which in one dimension corresponds to
pinning together the chiral right and left moving modes that without a mass are independent.
For this reason any perturbatively limited expansion of
correlators in terms of the mass term is meaningless.

We highlight the nontrivial effect of the mass term further by demonstrating that in a transient
regime it acts indeed as an effective temperature and defines the evolution in the
sense of a GGE. In the regime of spacetime distances shorter than the
length scale imposed by the mass, quantum fluctuations are large enough such that the system behaves as
being effectively massless. However, these excitations arise out of a ground state in which the right and left
moving modes are pinned together and maximally entangled. In the perturbation about this state the mass term
then enters as an effective temperature for these modes which, due to the large fluctuations, are seemingly
independent and massless otherwise.
GGEs are usually found in the stationary limit of states that cannot thermalize.
The situation is different here because we do not quench the ground state but instead consider
excitations out of it. Remarkably, a GGE behavior still appears but in the
transient and not in the stationary regime. We demonstrate that this is a robust and not
spurious feature by showing that it persists at any temperature if the effective temperature is subjected to
further mass and temperature corrections.

In the final part of this paper we propose approximations to the exact solutions that are
inspired by the insights gained from the preceding analysis. The exact solutions take the form of
Bessel functions that may be difficult to use if further nontrivial manipulations are required.
For instance, in effective many-body theories such as bosonization and the Luttinger liquid
description \cite{Haldane1981,vonDelft1998,Gogolin1998,Giamarchi2003,Shankar2017},
fermionic propagators are
obtained by exponentiating the bosonic propagators, and further Fourier transforming or similar
may be necessary. If the aim is not to just acquire numerical results
but to obtain a flexible working formula, an approximation that is treatable but maintains
the qualitative changes caused by a finite mass term is necessary.
To this end we introduce an approximation,
the profile function, that is expressed in terms of elementary functions but reproduces the
global profile of the exact solution in the entire spacetime. Furthermore it has parameters
that can be tuned to match particularly special regions of interest,
for instance the asymptotic limits.

Throughout the text we attempt to emphasize the physical relevance of the results and to point
out for what situation the latter can become useful. As we primarily provide the tools on which applications
can build within the broad range from the Planck scale to condensed matter, we believe that providing
such connections is more useful than focusing on one specific application as an illustration. The latter
may indeed risk to be either too niche or too involved and specific for the general messages we
intend to transmit with this paper.

The plan of the paper is as follows: In Sec.~\ref{sec:model} we define the model for the Klein-Gordon
theory and the various correlation functions we analyze. The exact solutions are derived in Sec.~\ref{sec:analytic},
at zero and finite temperature, the analytic structure is discussed, and the appropriate regularization is
introduced. In Sec.~\ref{sec:massless} we take the massless limit. The high temperature limit is
considered in Sec.~\ref{sec:high}, and the appearance of the transient generalized Gibbs ensemble in
Sec.~\ref{sec:GGE}. Section \ref{sec:profile} contains the profile function approximation. A discussion
and conclusion is provided in Sec.~\ref{sec:conclusions}. The appendices contain technical details
and proofs supporting the main text.


\section{Model and correlation functions}
\label{sec:model}

We consider a real scalar field $\phi$ spanning a $1+1$ dimensional space, with time coordinate $t$ and spatial
coordinate $x$, subject to the Klein-Gordon action
\begin{equation} \label{eq:S}
	S = \frac{\hbar}{2\pi v} \int dt\, dx\, \phi(t,x)
	\bigl[ \partial_t^2 - v^2 (\partial_x^2 + \Delta^2) \bigr]
	\phi(t,x),
\end{equation}
where $\hbar$ is the reduced Planck constant, $v$ is the velocity of the field, referred to as the light velocity
even if generally it may have a different origin within an effective theory, and
$\Delta$ is the mass term, expressed in units of inverse length and related to the true mass $m$ via the relation $m = \Delta \hbar / v$.
The field $\phi(t,x)$ is chosen dimensionless and real. In the discussion we will keep full units for clarity.

The action is Lorentz invariant, as is most directly seen by rewriting the derivatives in the integrand in the standard
notation $g^{\mu\nu} \partial_{\mu}\partial_{\nu}$, where $g^{\mu\nu} = \mathrm{diag}(1,-1)$ and $\mu,\nu \in \{0,1\}$
label $x^0 = vt$ and $x^1 = x$.
The Lorentz invariance is inherited by correlation functions at zero temperature, but it breaks down if the system is
coupled to a nonrelativistic thermal reservoir. Both situations will be discussed, and to this end we find it more
convenient to use directly the variables $t$ and $x$ instead of the notation $x^\mu$.

We aim to provide explicit analytic expressions for two-point correlations functions, whose building blocks are the
Wightman functions
\begin{align}
	W(t,x) &= \mean{\phi(t,x) \phi(0,0)},
\label{eq:W_tilde_definition}
\\
	\tilde{W}(t,x) &= \mean{\phi(0,0) \phi(t,x)},
\label{eq:W_definition}
\end{align}
which are equivalent to the also often used greater and lesser Green's functions, $W(t,x) = i \hbar G^>(t,x)$ and
$\tilde{W} = i \hbar G^<(t,x)$, respectively. From space and time translational invariance it follows that
$\tilde{W}(t,x) = W(-t,-x)$ and from the reality of the fields $\tilde{W}(t,x) = [W(t,x)]^*$.

The averages, $\mean{\dots}$, in Eqs.~\eqref{eq:W_definition} and \eqref{eq:W_tilde_definition} are taken either
over the ground state or over the
thermal ground state if the field is in contact with a thermal reservoir of temperature $T$.
In the latter case this takes the form of a thermal Gibbs ensemble, with operator expectations given by
\begin{equation}\label{eq:Thermal_avg}
	\mean{\dots} = \frac{1}{Z} \Tr\bigl\{ e^{- \beta H} \dots \bigr\}.
\end{equation}
Here $H$ is the Hamiltonian associated with the action, Eq.~\eqref{eq:S}, and $\beta = 1/k_B T$.
The chemical potential is set to zero throughout.
The trace $\Tr$ is taken over a full set of states and $Z = \Tr\{e^{-\beta H}\}$ is the partition function.
This explicit evaluation of thermal averages is not necessary though and below we will use instead standard relations
between the various correlation functions with products of spectral density and the Bose-Einstein distribution function.
In the limit $T \to 0$ the conventional ground state average is recovered.
We must emphasize that considering such a thermal reservoir is incompatible with Lorentz invariance because the Boltzmann
factors $e^{-\beta H}$ couple through $H$ only to energy $E$ but not to momentum. The existence of such a thermal reservoir is
therefore meaningful primarily if Eq.~\eqref{eq:S} describes an emergent action of a non-relativistic condensed matter system.
An effective temperature may arise also through the Unruh effect in an accelerating system \cite{Crispino2008}, but this is a
situation we will not address further.

Physical correlators or response functions in most cases do not rely on $W$ alone but on combinations of Eqs.~\eqref{eq:W_definition}
and \eqref{eq:W_tilde_definition} which can, through cancellations of various terms, lead to distinct properties.
We will thus investigate also the retarded and advanced Green's functions
\begin{align}
	G^r(t,x) &= -i \hbar^{-1} \theta(t) \mean{[\phi(t,x),\phi(0,0)]},
\label{eq:Gr}
\\
	G^a(t,x) &= i \hbar^{-1} \theta(-t) \mean{[\phi(t,x),\phi(0,0)]},
\label{eq:Ga}
\end{align}
where $\theta(t)$ is the unit step function, with $\theta(t) = 1$ for $t>0$, with $\theta(t) = 0$ for $t<0$, and with $\theta(0)$ undefined.
The reality of the scalar field permits the further simplification
\begin{align}
	G^r(t,x) &= 2 \hbar^{-1} \theta(t) \Imag[W(t,x)],
\label{eq:Gr2}
\\
	G^a(t,x) &= -2\hbar^{-1} \theta(-t) \Imag[W(t,x)].
\label{eq:Ga2}
\end{align}
We also consider the time-ordered, or Feynman, propagator, defined by
\begin{align}
	G^t(t,x) &= -i \hbar^{-1} \theta(t) \mean{\phi(t,x) \phi(0,0)}
	\nonumber\\
	&\quad
	-i \hbar^{-1} \theta(-t) \mean{\phi(0,0)\phi(t,x)}
	\nonumber\\
	&=
	-i \hbar^{-1} \bigl[ \theta(t) W(t,x) + \theta(-t) \tilde{W}(t,x) \bigr].
\label{eq:Gt}
\end{align}
Finally, we introduce
\begin{equation}
	G(t,x) = W(t,x)-W(0,0),
\end{equation}
which appears in averages over exponentiated fields such as $\mean{e^{i\phi(t,x)- i\phi(0,0)}} = e^{G(t,x)}$, which is a
common appearance in the context of bosonization and the Luttinger liquid \cite{Haldane1981,vonDelft1998,Gogolin1998,Giamarchi2003,Shankar2017}
and hence central for condensed matter applications.
In the massless case, the subtraction of $W(0,0)$ is necessary to regularize infrared divergences. For the case of a
nonzero mass this regularization requires some extra care, and will be the subject of the discussion in Sec.~\ref{sec:regularization}.

The relations between the different correlation functions are best seen in $(\omega,k)$ space, which we connect to
the $(t,x)$ representation using the Fourier transform conventions
\begin{align}
	f(\omega) &= \int_{-\infty}^\infty dt \, e^{i \omega t} f(t),
\\
	f(k) &= \int_{-\infty}^\infty dx \, e^{-i k x} f(x),
\end{align}
for any function $f$.
The central quantity is the spectral density $A(\omega,k)$, defined by
\begin{equation} \label{eq:A_def}
	A(\omega,k) = - \frac{1}{\pi} \Imag G^r(\omega,k).
\end{equation}
Using standard identities that for completeness are reproduced in \ref{sec:identities}, this function allows us to write
\begin{align}
	W(\omega,k) &= \frac{2\pi\hbar}{1-e^{-\hbar\beta \omega}} A(\omega,k),
\label{eq:fluct_diss}
\\
	\tilde{W}(\omega,k) &= \frac{2\pi\hbar e^{-\hbar\beta \omega}}{1-e^{-\hbar\beta \omega}} A(\omega,k).
\label{eq:fluct_diss_1}
\end{align}
Taken together, these relations produce the fluctuation-dissipation theorem
\begin{equation} \label{eq:fluct_diss_2}
	W(\omega,k) + \tilde{W}(\omega,k) = 2\pi \hbar \coth\left(\frac{\hbar\beta \omega}{2} \right)A(\omega,k).
\end{equation}
As shown in \ref{sec:identities} the spectral function $A(\omega,k)$ is temperature independent,
and hence maintains the Lorentz invariance of the action. In the zero temperature limit $\beta \to \infty$,
Eqs.~\eqref{eq:fluct_diss} and \eqref{eq:fluct_diss_1} yield
\begin{align}
	W(\omega,k) &\to 2\pi\hbar\theta(\omega) A(\omega,k),
\label{eq:fluct_diss_limit}
\\
	\tilde{W}(\omega,k) &\to -2\pi\hbar\theta(-\omega) A(\omega,k).
\label{eq:fluct_diss_limit_1}
\end{align}
Taking the Fourier transform of Eqs.~\eqref{eq:fluct_diss} and \eqref{eq:fluct_diss_1} carries an advantage over explicit
evaluations of Eq.~\eqref{eq:Thermal_avg} in the form of separating the fundamental Lorentz-invariant contribution to $W$
and $\tilde{W}$ from the non-relativistic effect of the thermal reservoir.


\section{Analytic solutions}
\label{sec:analytic}

The primary interest in the two-point correlation functions usually does not lie in themselves
directly but in their use as building blocks for observables and response functions, often in expansions
of a many-body theory. As such their analytic properties have a direct impact on the physical characteristics.
We therefore start with a thorough derivation of the possible analytic solutions and provide an extended
discussion of their properties.

\subsection{Integral representation}

The starting point for the determination of $W(t,x)$ is the spectral function in $(\omega,k)$, given by
\begin{equation}\label{eq:A(w,k)_gapped}
	A(\omega,k)
	= \frac{\pi v}{2\hbar\omega_k}\bigl[\delta(\omega-\omega_k)-\delta(\omega+\omega_k)\bigr],
\end{equation}
with
\begin{equation}\label{eq:omega_k_gapped}
	\omega_k = v \sqrt{ k^2 + \Delta^2}.
\end{equation}
A derivation of this standard result is provided in \ref{sec:spectral}. Through Eq.~\eqref{eq:fluct_diss} it follows
\begin{equation}
	W(\omega,k)
	= \frac{2\pi}{1-e^{-\beta\hbar\omega}}\frac{\pi v}{2\omega_k}\bigl[\delta(\omega-\omega_k)-\delta(\omega+\omega_k)\bigr],
\end{equation}
which needs to be Fourier transformed twice to $(t,x)$ space. The Fourier transform to time $t$ is immediate,
\begin{align}
	&W(t,k)
	= \int_{-\infty}^\infty \frac{d\omega}{2\pi} e^{-i\omega t} W(\omega,k)
\nonumber\\
	&=
	\frac{\pi v}{2\omega_k}
	\left(
		\frac{e^{-i \omega_k t}}{1-e^{-\beta\hbar\omega_k}}
		-
		\frac{e^{i \omega_k t}}{1-e^{\beta\hbar\omega_k}}
	\right)
\nonumber\\
	&=
	\frac{\pi v}{2\omega_k}
	\bigl[
		\cos(t \omega_k) \coth(\beta \hbar \omega_k/2) - i \sin(t \omega_k)
	\bigr],
\end{align}
and $W(t,x)$ is therefore expressed by the Fourier integral
\begin{align}
	&W(t,x)
	=
	\int_{-\infty}^\infty \frac{dk}{2\pi} e^{i k x} \, W(t,k)
\nonumber\\
	&=
	\frac{v}{4}
	\int_{-\infty}^\infty \frac{dk}{\omega_k} e^{i k x}
	\bigl[
		\cos(t \omega_k) \coth(\beta \hbar \omega_k/2) - i \sin(t \omega_k)
	\bigr].
\nonumber\\
\label{eq:W(t,x)_int}
\end{align}
We shall first evaluate this integral in the zero-temperature limit, which then informs a systematic extension to finite temperatures.


\subsection{Zero temperature and Lorentz invariance}

The zero temperature limit of $W(t,x)$ is characterized by $\beta \to \infty$ and $\coth(\beta \hbar \omega_k/2) \to 1$ in
Eq.~\eqref{eq:W(t,x)_int}. If we write $W_0(t,x)$ to denote this limit, the integral to evaluate becomes
\begin{equation} \label{eq:W0_int}
	W_0(t,x)
	=
	\frac{v}{4}
	\int_{-\infty}^\infty \frac{dk}{\omega_k} e^{i k x - i t \omega_k}.
\end{equation}
This expression must be Lorentz invariant. This can indeed be seen by the change of variable $k = \Delta \sinh(\varphi)$,
from which we have $\omega_k = v \Delta \cosh(\varphi)$ and
\begin{equation}
	W_0(t,x)
	=
	\frac{1}{4}
	\int_{-\infty}^\infty d\varphi \, e^{-i \Delta [ \cosh(\varphi) v t- \sinh(\varphi) x]}.
\end{equation}
The expression in the exponent is the time component of an orthochronous Lorentz transformation $(vt,x) \to (vt',x')$ with
rapidity $\varphi$, such that $vt' = \cosh(\varphi) v t - \sinh(\varphi) x$. Since $d\varphi$ represents the Haar measure of the
orthochronous Lorentz group, as $d(\varphi+\vartheta) = d\varphi$ for any constant $\vartheta$,
the value of $W_0(t,x)$ is the same along the curves of constant distance $s^2 = v^2t^2-x^2$ and fixed $\sign(t)$,
thus proving the Lorentz invariance.

Further evaluation of this integral depends on whether $(t,x)$ is time-like or space-like.
For space-like coordinates $|vt/x|<1$ and we can set $vt = |s| \sinh(\vartheta)$ and $x = \sign(x) |s| \cosh(\vartheta)$
(cf.\ also Refs.~\cite{Pais1950,Feinberg1963}). This leads to
\begin{align}
	&\cosh(\varphi) vt - \sinh(\varphi) x
\nonumber\\
	&=
	|s| \bigl[
		\cosh(\varphi) \sinh(\vartheta) - \sign(x) \sinh(\varphi) \cosh(\vartheta)
	\bigr]
\nonumber\\
	&=
	|s| \sinh\bigl(\varphi - \sign(x) \vartheta\bigr).
\end{align}
Since the integration measure is independent of any constant shift, we find
\begin{align}
	W_0(t,x)
	&=
	\frac{1}{4}
	\int_{-\infty}^\infty d\varphi \, e^{-i \Delta |s| \sinh(\varphi)}
\nonumber\\
	&=
	\frac{1}{2}
	\int_0^\infty d\varphi \, \cos\bigl(\Delta |s| \sinh(\varphi)\bigr).
\label{eq:W0_varphi}
\end{align}
This integral is of a standard type \cite[Eq.~\href{http://dlmf.nist.gov/10.32.E6}{(10.32.6)}]{DLMF} and results in
\begin{equation} \label{eq:W0_t>0}
	W_0(t,x) = \frac{1}{2} K_0\bigl(\Delta \sqrt{x^2-v^2t^2}\bigr),
\end{equation}
for $x^2 > v^2 t^2$, where $K_0$ is the modified Bessel function.

For time-like coordinates $|vt/x| > 1$ and we define $\vartheta$ now through
$vt = \sign(t)|s| \cosh(\vartheta)$ and $x = -\sign(t) |s| \sinh(\vartheta)$.
The inclusion of $\sign(t)$ is necessary so that the following addition theorem can be used:
\begin{align}
	&\cosh(\varphi) vt - \sinh(\varphi) x
\nonumber\\
	&=
	\sign(t) |s| \bigl[
		\cosh(\varphi) \cosh(\vartheta) + \sinh(\varphi) \sinh(\vartheta)
	\bigr]
\nonumber\\
	&=
	\sign(t) |s| \cosh\bigl(\varphi + \vartheta\bigr).
\end{align}
The shift by $\vartheta$ is once again of no importance for the integration and we obtain
\begin{align}
	W_0(t,x)
	=
	\frac{1}{4}
	\int_{-\infty}^\infty d\varphi \, e^{-i \sign(t) \Delta |s| \cosh(\varphi)}.
\end{align}
This is another standard integral \cite[Eq.~\href{http://dlmf.nist.gov/10.9.E8}{(10.9.8)}]{DLMF} and solves to
\begin{align}
	&W_0(t,x)
\nonumber\\
	&=
	-\frac{\pi}{4}
	\Bigl[
		Y_0\bigl(\Delta\sqrt{v^2t^2-x^2}\bigr)
		+ i \sign(t)
		J_0\bigl(\Delta\sqrt{v^2t^2-x^2}\bigr)
	\Bigr]
 \nonumber\\
	&=
	\begin{cases}
		-\frac{i\pi}{4}
		H_0^{(2)}\bigl(\Delta\sqrt{v^2t^2-x^2}\bigr), &{t>0},
	\\
		\frac{i\pi}{4}
		H_0^{(1)}\bigl(\Delta\sqrt{v^2t^2-x^2}\bigr), &{t<0},
	\end{cases}
\label{eq:W0_t<0}
\end{align}
where $J_0$ and $Y_0$ are Bessel functions of the first and second kind, respectively,
and $H^{(1)}_0, H^{(2)}_0$ are Hankel functions.

Both space-like and time-like expressions can be unified through the connection formulas
$K_0(z e^{-i \pi/2}) = \frac{i\pi}{2} H_0^{(1)}(z)$ and
$K_0(z e^{i\pi/2}) = -\frac{i\pi}{2}H_0^{(2)}(z)$ for $z>0$
\cite[Eq.~\href{http://dlmf.nist.gov/10.27.E8}{(10.27.8)}]{DLMF}.
For time-like coordinates we have $\sqrt{x^2-(vt-i\eta)^2} = \sqrt{v^2t^2-x^2} e^{i \sign(t) \pi/2}$
for infinitesimal $\eta >0$, if we choose here (as henceforth) the principal branch of the square root.
It thus follows that
\begin{equation} \label{eq:W0_final}
	W_0(t,x) = \frac{1}{2} K_0\bigl(\Delta \sqrt{x^2-(vt-i\eta)^2}\bigr).
\end{equation}
The latter equation provides a single, exact, analytic formula valid for all $t,x$ off the light
cone, $s^2 \neq 0$, since the shift $\eta$ is of no consequence in the space-like region.
As $\eta$ is infinitesimal, the Lorentz invariance of $W_0(t,x)$ remains indeed valid everywhere
except on the light cone, where the solution diverges and requires a separate discussion that will
be provided in Sec.~\ref{sec:regularization}.

\begin{figure}
	\centering
	\includegraphics[width=\columnwidth]{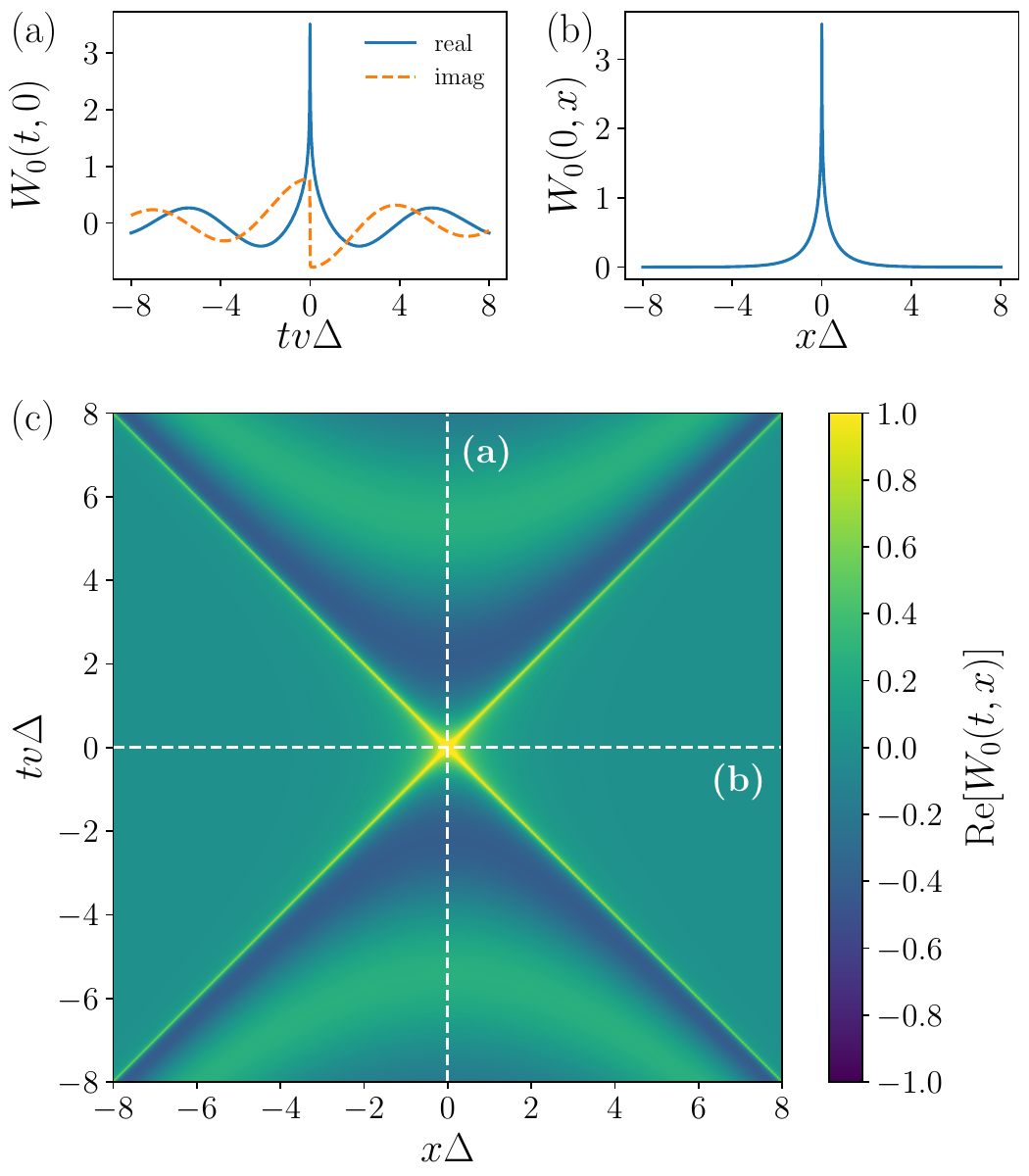}
	\caption{\label{fig:W0}%
		Illustration of the Wightman function $W_0(t,x)$ as a function of time $t$ and position $x$.
		Panel (a) shows the time-like behavior at $x=0$ and $t>0$, which displays after an initial
		fast drop an algebraic decay with oscillating real and imaginary parts. The divergence of
		$\Real[W_0(t,0)]$ and the discontinuity of $\Imag[W_0(t,0)]$ at the origin are regularized
		by the cutoff $\alpha$ (see Sec.~\ref{sec:regularization}), chosen to be $\alpha = 0.001/\Delta$.
		Panel (b) shows the space-like behavior at $t=0$ that has an exponential decay
		for larger $x$ and is purely real.
		Panel (c) displays the wider $(t,x)$ behavior of $\Real[W_0(t,x)]$.
		A similar pattern would arise for $\Imag[W_0(t,x)]$ in the time-like regions but in the
		space-like regions $\Imag[W_0(t,x)]=0$.
		The Lorentz invariance of the function is well visible by the hyperbolic shape of
		the fringes in the time-like regions.
		The color axis is truncated at values $\pm 1$ for better visibility of the features.
		The white dashed lines mark the cuts displayed in panels (a) and (b).
	}
\end{figure}

An illustration of the distinct behavior for time-like and space-like arguments $(t,x)$ is
shown in Fig.~\ref{fig:W0}. Notable is that $W_0(t,x)$ is real for space-like arguments, whereas for
time-like arguments it has oscillating real and imaginary parts. Further features of $W_0$ and their
underlying physics will be discussed in detail later in the paper.


\subsection{Temperature dependence}
\label{subsec:Temp_Dependence}
The coupling to a nonrelativistic thermal reservoir breaks the Lorentz invariance.
Turning back to Eq.~\eqref{eq:W(t,x)_int} for arbitrary inverse temperature $\beta$, we can use the expansion
$\coth(\beta\hbar\omega_k/2) = 1 + 2 \sum_{n\ge 1} e^{-n \beta \hbar \omega_k}$
to write
\begin{equation}
	W(t,x) = W_0(t,x) + \sum_{n \ge 1} W_n(t,x),
\end{equation}
where
\begin{equation} \label{eq:W_n_1}
	W_n(t,x) = \frac{v}{2} \int_{-\infty}^\infty \frac{dk}{\omega_k} e^{i k x} \cos(t \omega_k) e^{- n \beta \hbar \omega_k}.
\end{equation}
for $n \ge 1$. Letting $k = \Delta \sinh(\varphi)$ and $\sqrt{k^2+\Delta^2} = \Delta \cosh(\varphi)$
this can be written in a similar way as for $W_0$ before as
\begin{align}
	&W_n(t,x)
	=
	\frac{1}{4} \int_{-\infty}^\infty d\varphi
	\Bigl[
		e^{ i \Delta x \sinh(\varphi) + \Delta v (it - n \beta \hbar) \cosh(\varphi)}
\nonumber\\
	&\qquad
	+
		e^{ i \Delta x \sinh(\varphi) + \Delta v (-it - n \beta \hbar) \cosh(\varphi)}
	\Bigr].
\label{eq:W_n_2}
\end{align}
These integrals resemble those done for $W_0$ but with the difference that the exponent is
no longer purely imaginary and causes an additional exponential convergence. It remains possible to express the $W_n$ in terms
of the $K_0$ Bessel function though. To see this we rewrite the $(t,x)$ variables such that the exponential convergence remains
evident and can be used to control deformations of the complex integration contour,
\begin{align}
	&W_n(t,x)
	=
\nonumber\\
	&\frac{1}{4}
	\int_{-\infty}^\infty d\varphi \Bigl[
		e^{- \xi_+^n \cosh(\varphi+\vartheta_+^n)}
		+
		e^{- \xi_-^n \cosh(\varphi+\vartheta_-^n)}
	\Bigr],
\end{align}
with $\xi_\pm^n = \Delta \sqrt{x^2 - v^2(t\pm i n \beta \hbar)^2}$
and
$\tanh(\vartheta_\pm^n) = (-ix) / [-v(\pm i t - n \beta\hbar)]$.
We change the integration variable $\varphi \to \varphi-\vartheta_\pm^n$ such that
\begin{align}
	&\int_{-\infty}^\infty d\varphi \, e^{- \xi_\pm^n \cosh(\varphi+\vartheta_\pm^n)}
\nonumber\\
	&=
	\int_{-\infty+i\Imag[\vartheta_\pm^n]}^{\infty+i\Imag[\vartheta_\pm^n]} d\varphi \, e^{- \xi_\pm^n \cosh(\varphi)},
\label{eq:shifted_contour}
\end{align}
where the principal branches of $\tanh^{-1}$ and the square root are chosen, in addition to the positive solution $\Real\xi_\pm^n>0$.
By construction, as proved in detail in \ref{sec:contour_deformation_proof}, $\Real[\xi_\pm^n \cosh(\varphi)]>0$
for all $\varphi$ on the integration contour. Therefore, the contour can be smoothly
deformed to run over the real axis, $\varphi \in \mathbb{R}$, while maintaining strict convergence and analyticity of the integrand.
As a result we have
\begin{align}
	&W_n(t,x)
	=
	\frac{1}{4}
	\int_{-\infty}^\infty d\varphi \Bigl[
		e^{- \xi_+^n \cosh(\varphi)}
		+
		e^{- \xi_-^n \cosh(\varphi)}
	\Bigr].
\label{eq:shifted_contour_1}
\end{align}
The integrals are thus brought to yet another integral definition of the $K_0$ Bessel function with
complex argument \cite[Eq.~\href{http://dlmf.nist.gov/10.32.E9}{(10.32.9)}]{DLMF},
\begin{equation}
	K_0(z) = \int_0^\infty d\vartheta \, e^{-z \cosh(\varphi)},
\end{equation}
for $\Real z >0$, such that
\begin{align}
	&W_n(t,x)
	=
	\frac{1}{2}
	\bigl[
		K_0(\xi_+^n) + K_0(\xi_-^n)
	\bigr].
\label{eq:W_n_post_contour_deform}
\end{align}
Noting that $\xi_-^{n} = \xi_+^{-n}$ we finally obtain
\begin{align}
	W(t,x)
	&= \frac{1}{2} K_0\bigl(\Delta \sqrt{x^2-v^2(t-i\eta)^2}\bigr)
\nonumber\\
	&+\frac{1}{2} \sum_{n \neq 0} K_0\bigl(\Delta \sqrt{x^2-v^2(t-i n \beta \hbar)^2}\bigr).
\label{eq:W_full}
\end{align}
It should be stressed that the square roots must be chosen such that their real parts are positive and their
imaginary parts correspond to the principal branch which connects smoothly to $\Imag\sqrt{(\dots)} \to 0$.
We note furthermore that Eq.~\eqref{eq:W_full} is invariant under imaginary time
shifts (inverse of Matsubara frequencies) $t \to t + i n \beta \hbar$ for integer $n$, as required for bosonic correlators,
as long as we keep an infinitesimal $i\eta$ in the term that is shifted to $n=0$.
This additionally corroborates the possibility and necessity to deform the $\varphi$ integration contour
as described above.
Furthermore, it is clear that the temperature dependent correction to $W_0$ is real because $\xi_-^n = \xi_+^{-n}=(\xi_+^n)^*$.
We also note that unlike $W_0$, each $W_n$ remains finite on the light cone.
The behavior at high and low temperatures shall be discussed in more detail in Secs.~\ref{sec:high} and \ref{sec:GGE}.


\subsection{Retarded, advanced, and time ordered functions}

The retarded function is obtained from Eq.~\eqref{eq:Gr2} by the imaginary part of
$W(t,x)$. Therefore, it is only non-zero inside the light cone at $t>0$ and
temperature independent since the temperature corrections are real. The result is
\begin{align}
	G^r(t,x) = -\theta(t) \theta(v^2 t^2-x^2)
	\frac{\pi}{2\hbar}
	J_0\bigl(\Delta \sqrt{v^2 t^2-x^2}\bigr).
\label{eq:Gr_final}
\end{align}
The temperature independence reflects also that the thermal reservoir cannot affect causality,
and $G^r$ remains nonzero only inside the light cone at positive times, in addition to preserving
Lorentz invariance.
The advanced function, Eq.~\eqref{eq:Ga2}, is accordingly
\begin{align}
	G^a(t,x)
	= -\theta(-t) \theta(v^2 t^2-x^2)
	\frac{\pi}{2\hbar}
	J_0\bigl(\Delta \sqrt{v^2 t^2-x^2}\bigr),
\label{eq:Ga_final}
\end{align}
which is nonzero only inside the light cone at $t<0$.

From Eq.~\eqref{eq:Gt} we see that the time dependent correlator depends on $W(t,x)$
for $t>0$ and on $[W(t,x)]^*$ for $t<0$. We can therefore write
\begin{align}
	G^t(t,x) = &-\frac{i}{2\hbar}K_0\bigl(\Delta \sqrt{x^2-v^2(t-i\eta \sign(t))^2}\bigr)
\nonumber\\
	&- \frac{i}{2\hbar} \sum_{n \neq 0} K_0\bigl(\Delta \sqrt{x^2-v^2(t-i n \beta \hbar)^2}\bigr),
\end{align}
noting that the temperature dependent sum does not change for the different operator orderings.

On the other hand, the function $G(t,x) = W(t,x)-W(0,0)$ is divergent due to the logarithmic singularity
of $K_0$ at the origin, leading to the necessity of a regularization scheme as discussed in the following.


\subsection{Regularization of singularities on the light cone}
\label{sec:regularization}

When approaching the light cone, $s^2 = v^2 t^2 - x^2 \to 0$, the temperature dependent parts of
$W(t,x)$ remain finite but the temperature independent part, $W_0(t,x)$, has through the
asymptotics of the Bessel function \cite[Eq.~\href{http://dlmf.nist.gov/10.31.E2}{(10.31.2)}]{DLMF} the
divergent behavior
\begin{equation} \label{eq:W0_divergent}
	W_0(t,x) \sim -\frac{1}{2} \ln\bigl(\Delta \sqrt{x^2-(vt-i\eta)^2}\bigr).
\end{equation}
Equation \eqref{eq:W0_int} shows that this divergence is ultraviolet, as the integrand behaves
as $1/k$ for $|k| \gg \Delta$ and produces a logarithmic divergence that on the light cone is no longer suppressed by
the oscillating exponential.
Such a divergence is an intrinsic feature of many quantum field theories, and a large number of regularization methods have
been devised. Common types are the Pauli-Villars \cite{Pauli1949}, dimensional \cite{Bollini1972,tHooft1972,Bollini1996}, and
analytic \cite{Speer1968,Speer1971} regularizations, if we focus on regularization of the continuum theory
and exclude discretizations such as lattice regularization as justified further below. These regularization
schemes are constructed to either subtract or to suppress the divergent terms and generally assume the regularization to be
applied on the divergent integration. We have to notice, however, that off the light cone the theory has through the
oscillating exponentials a natural regularization and converges, producing the analytic result of the Bessel function.
Any further regularization scheme on the light cone needs to be consistent with this result and needs to have no
influence where it is not required.

Practically there are two parameters, the spacetime distance $s$ and a regularization parameter, say $\alpha$, such
that $\alpha \to 0$ represents the unregularized theory. The limits $s \to 0$ and $\alpha \to 0$ generally do not commute.
Therefore, a regularization only on the light cone or at $x=vt=0$ as often considered would remain discontinuous from
the remaining $W(t,x)$ and consequently must be considered as being of limited use.

As an example, we consider the Pauli-Villars regularization, which consists in subtracting a correlator of the same form
but with a different mass, $\Delta' = 1/\alpha$, that is
at the end sent to infinity. If we demand that this subtraction holds for all coordinates
then we would need to replace $W_0(t,x)$ by
\begin{equation} \label{eq:PV}
	W^\text{PV}_0(t,x) = \lim_{\Delta'\to \infty} \frac{1}{2} \Bigl[
		K_0\bigl(\Delta \sqrt{-s^2}\bigr)
		-
		K_0\bigl(\Delta' \sqrt{-s^2}\bigr)
	\Bigr],
\end{equation}
with $s^2 = v^2t^2-x^2$. As long as $s\neq 0$ the limit suppresses the second term
and we recover the unregularized $W_0$ of Eq.~\eqref{eq:W0_final}. Although consistent for most $(t,x)$ this does not resolve
the divergence on the light cone. Taking instead the $s \to 0$ limit first leads through Eq.~\eqref{eq:W0_divergent} to
\begin{equation}
	W^\text{PV}_0(t,x) \sim \lim_{\Delta'\to \infty}
		\frac{1}{4}
		\ln\bigl(\Delta'/\Delta\bigr).
\end{equation}
This confirms that the divergence is logarithmic, but is decoupled from $(t,x)$
and thus does not capture how the divergence builds up from a physical approach to the light cone.

As a further example, analytic continuation schemes such as those discussed in Refs.~\cite{Speer1968,Speer1971}
modify the power of the square root to some $1/2 + \alpha$, with $\alpha>0$, in the denominator of the integrand
of Eq.~\eqref{eq:W0_int}, which leads to
\begin{equation} \label{eq:W_0^A}
	W_0^A(t,x) = \frac{1}{4} \int_{-\infty}^\infty \frac{dk \, \Delta^{2\alpha}}{(k^2+\Delta^2)^{\frac{1}{2}+\alpha}} e^{i k x - i \omega_k t},
\end{equation}
where $\Delta^{2\alpha}$ ensures correct dimensions. Due to the extra $(k^2+\Delta^2)^{-\alpha}$ factor Lorentz invariance is now
broken. Furthermore we are unable to evaluate this integral exactly, except when $t=0$. In this case we obtain once more a standard
integral for the modified Bessel function \cite[Eq.~\href{http://dlmf.nist.gov/10.32.E11}{(10.32.11)}]{DLMF}, such that
\begin{equation}
	W_0^A(0,x) = \frac{1}{2} (\Delta |x|)^{\alpha} \frac{\sqrt{\pi}}{2^\alpha \Gamma(\frac{1}{2}+\alpha)} K_\alpha(\Delta|x|),
\end{equation}
with $\Gamma$ Euler's Gamma function. We can discard $\sqrt{\pi}/2^\alpha \Gamma(\frac{1}{2}+\alpha) = 1 + \mathcal{O}(\alpha)$
and restore Lorentz invariance by the boost $|x| \to \zeta = \sqrt{x^2 - (vt-i\eta)^2}$, similar to a procedure used
in Ref.~\cite{Starykh1999}.
The result,
\begin{equation} \label{eq:W_0^A_final}
	W_0^A(t,x) = \frac{1}{2} (\Delta \zeta)^{\alpha}
	K_\alpha(\Delta \zeta),
\end{equation}
provides an appropriately regularized solution, yielding $W_0^A \sim \Gamma(\alpha) \sim 1/\alpha$ on the light cone and
analytic continuity through the light cone.

Nevertheless we shall opt against using this regularization. The primary  reason is that it is formalistic, using the terminology
introduced by Pauli and
Villars \cite{Pauli1949}, in that it is a mathematical regularization that is not set or inspired by a physical quantity.
While we do not consider this as a deficit per se, we find the formalistic structure difficult to reconcile with the extension to $T>0$
and the limit of a vanishing mass, $\Delta \to 0$.
For a nonzero temperature, the terms $W_n$ for $n \neq 0$ themselves do not require regularization and indeed the imaginary time shifts
$i n \beta \hbar$ guarantee full convergence. Yet to maintain full periodicity under the shifts $i n \beta \hbar$ it would be necessary
to extend these terms to expressions of the type $W_n^A = W_0^A(t + i n \beta \hbar,x) + W_0^A(t - i n \beta \hbar,x)$ that
mix the formalistic regularization $\alpha$
with the physical regularization $i n \beta\hbar$. For a vanishing mass, $\Delta \to 0$, the denominator would become $|k|^{1+2\alpha}$ and
hence for $\alpha > 0$ make the infrared divergence at $k \to 0$ even worse. Therefore, in the context of the wider physical embedding
of $W_0(t,x)$ considered here, the analytic continuation becomes too problematic so that we shall discard it.
Similar issues appear with dimensional regularization \cite{Bollini1972,tHooft1972,Bollini1996}.
If, however, none of these
situations is of importance, then Eq.~\eqref{eq:W_0^A_final} remains a perfectly valid regularization, and most of
the discussion to follow in the next sections at $T=0$ and $\Delta \neq 0$ would remain then applicable too.

Alternative approaches accept the singular behavior but diminish its influence by convolving it
with a measurement function, for instance from a finite sized detector \cite{Schlicht2004,Langlois2006} or by the application of
positive operator-valued measures \cite{Bednorz2023}. As such approaches are useful but do not address
head-on the analytic continuity of the correlators we shall here not consider them further.
For the same reason we do not consider lattice regularizations as they truncate also the
analytic structure of the correlators underlying much of our study.

We will instead employ a scheme that restricts the regularization to the immediate vicinity of the light cone while maintaining
continuity of $W_0(t,x)$ throughout. If we further impose that the analytic behavior remains unchanged,
then Eq.~\eqref{eq:W0_final} already provides the solution, obtained by replacing the infinitesimal
$\eta$ by a finite, small cutoff length $\alpha > 0$,
\begin{equation}
\label{eq:W0_alpha}
	W_0(t,x) = \frac{1}{2} K_0\bigl(\Delta \sqrt{ x^2 - (vt -i \alpha)^2}\bigr).
\end{equation}
We thus argue that the simplest regularization scheme, by an ultraviolet cutoff, is here the
most appropriate choice, and that among the many ways a cutoff can be introduced only
the shift of the time variable as in Eq.~\eqref{eq:W0_alpha} is consistent with the analytic
properties of the correlator, as well as with the physical limits of $T \neq 0$ and
$\Delta \to 0$, as shall be shown. Equation \eqref{eq:W0_alpha} therefore is the
regularization scheme we employ henceforth. We should note that the same result is obtained
by applying a broadening, or smearing, procedure directly on the fields $\phi$ \cite{Radovskaya2023}.

It must be noted though that this regularization also breaks the Lorentz invariance in the
narrow intervals of width proportional to $\alpha$ about the light cone.
For a condensed matter effective field theory this is acceptable because Lorentz invariance is an emergent and not a
fundamental property of the low energy physics. In such a case the cutoff $\alpha$ appears as a natural
regularization choice through the system's bandwidth or the microscopic lattice spacing, which in
addition has the advantage of being a ``realistic theory'', in Pauli and Villars' terminology \cite{Pauli1949},
in contrast to the ``formalistic theory'' mentioned earlier.
It is also worth mentioning that in a condensed matter system one often abandons the underlying microscopic
lattice theory although it is perfectly regularised, in order to capture the emergent universal behavior of
the continuum theory. The advantage of universality and, in fact the appearance of the discussed
analytic properties of the solutions, offsets the principally unnecessary reintroduction of
the cutoff $\alpha$. For this reason we do not consider lattice regularization in our treatment either.

For an elementary field theory it would be
desirable though to maintain Lorentz invariance. For the present case, as already hinted by the discussed difficulties
with the regularization schemes, this requirement principally cannot be satisfied in a continuous way.
This can be seen from Eq.~\eqref{eq:W0_t<0} in which
the dependence on $\sign(t)$ contradicts continuity through the point $(t=0,x=0)$. Lorentz invariance
furthermore would impose that $W_0(0,0)$ is equal to all $W_0(t,x)$ along the light cones for
positive and negative $t$, which again would
be in contradiction with the $\sign(t)$ in Eq.~\eqref{eq:W0_t<0}. With the chosen cutoff $\alpha$
the passage through the light cone instead depends on the argument
\begin{equation}
	x^2 - v^2 t^2 + \alpha^2 + 2 i v t \alpha \sim \alpha^2 + 2 i v t \alpha,
	\quad \text{as $s \to 0$},
\end{equation}
in which $\alpha^2$ cuts off the singularity and $2ivt\alpha$ at $t \to 0$ interpolates
the $\sign(t)$ function smoothly between $+1$ and $-1$ and vanishes at $t=0$.
These advantages compensate in our opinion the small error made on the Lorentz invariance
near the light cone.

As a consequence the retarded and advanced functions now would leak into the space-like
regions though by a distance $\sim \alpha$. However, as seen from Eqs.~\eqref{eq:Gr_final}
and \eqref{eq:Ga_final} these functions only depend on the Bessel function $J_0$ which
remains finite throughout as $J_0(0) = 1$.
Therefore a regularization of $G^r$ and $G^a$ is unnecessary, and the leakage into the
forbidden space-like region can be avoided.

It is furthermore instructive to see how this cutoff appears in the $k$ integral.
As it consists of an imaginary shift of $t$ we see that Eq.~\eqref{eq:W0_int} is regularized as
\begin{equation}\label{eq:W0_int_regularized}
	W_0(t,x)
	=
	\frac{v}{4}
	\int_{-\infty}^\infty \frac{dk}{\omega_k} e^{i k x - i t \omega_k} e^{- \alpha \omega_k/v},
\end{equation}
and that $\alpha$ represents a smooth cutoff at large energies. It is notable that this
cutoff couples to $\omega_k = v \sqrt{k^2+\Delta^2}$ and not directly to momentum $k$.
This is formally identical to the smearing procedure employed in Ref.~\cite{Radovskaya2023}.

To conclude this discussion, let us turn to the temperature dependent terms in $W(t,x)$.
Since they remain finite on the light cone they do not require regularization.
However, through the substitution $t \to t-i\alpha$ in $W_0$ the full result loses the
exact periodicity in the imaginary time direction by any shift $t \to t + i n \beta \hbar$
for integer $n$. To maintain this periodicity we need to shift
the time argument for all terms in the temperature expansion such that
\begin{align}
	W(t,x)
	&= \tilde{W}(-t,-x)
	= \frac{1}{2} K_0\bigl(\Delta \sqrt{ x^2 - (vt -i \alpha)^2}\bigr)
\nonumber\\
	&+ \frac{1}{2}\sum_{n \neq 0} K_0\bigl(\Delta \sqrt{ x^2 - (vt - i \alpha - i n v \beta \hbar)^2}\bigr).
\end{align}
Through this shift the terms with opposite $+n$ and $-n$ no longer are complex conjugate to each
other and the temperature dependent contribution no longer is purely real. Taking the imaginary part
of $W(t,x)$ to produce retarded and advanced functions would then cause these functions to have a
temperature dependence and nonzero values in the space-like region near the light cone.
Such an approach should therefore be avoided and instead the regularization free, and temperature
independent, expressions of
Eqs.~\eqref{eq:Gr_final} and \eqref{eq:Ga_final} should be used throughout.

It should moreover be noted that the substitution $t \to t - i \alpha$ is incompatible with the
multiplication by $e^{- \alpha \omega_k/v}$ in the integrals defining $W_n$. As seen in Eq.~\eqref{eq:W_n_2}
an extra factor $e^{- \alpha \omega_k/v}$ inside the integral would instead lead to $t + i \alpha$ and $t - i\alpha$ shifts
for the two terms of the integrand, respectively. This would preserve the reality of $W_n$ but alter the analytic
properties of the full correlator. For the present discussion we opt to preserve the analytic properties, but
if the analytic properties are of less concern then a choice preserving reality instead or the omission of
$\alpha$ for the temperature terms could be made as well. We shall come back again to this topic when discussing the
massless correlators below.


\subsection{Summary of the analytic results}
\label{sec:summary_analytic}

To summarize, the different correlation functions have the explicit analytic expressions
\begin{align}
	&W(t,x)
	= \tilde{W}(-t,-x)
	= \frac{1}{2} K_0\bigl(\Delta \sqrt{ x^2 - (vt -i \alpha)^2}\bigr)
\nonumber\\
	&+ \frac{1}{2}\sum_{n \neq 0} K_0\bigl(\Delta \sqrt{ x^2 - (vt -i\alpha -i n v \beta \hbar)^2}\bigr),
\\
	&G^r(t,x)
	= -\theta(t) \theta(v^2 t^2-x^2)
	\frac{\pi}{2\hbar}
	J_0\bigl(\Delta \sqrt{v^2 t^2-x^2}\bigr),
\\
	&G^a(t,x)
	= -\theta(-t) \theta(v^2 t^2-x^2)
	\frac{\pi}{2\hbar}
	J_0\bigl(\Delta \sqrt{v^2 t^2-x^2}\bigr),
\\
	&G^t(t,x)
	= -\frac{i}{2\hbar} K_0\bigl(\Delta \sqrt{ x^2 - (vt -i \alpha \sign(t))^2}\bigr)
\nonumber\\
	&- \frac{i}{2\hbar}\sum_{n \neq 0} K_0\bigl(\Delta \sqrt{ x^2 - (vt - i\alpha \sign(t) - i n v \beta \hbar)^2}\bigr),
\\ \label{eq:G_gapped_full}
	&G(t,x)
	= W(t,x)-W(0,0)
\nonumber\\
	&= \frac{1}{2} \left[
		K_0\bigl(\Delta \sqrt{ x^2 - (vt -i \alpha)^2}\bigr)
		-
		K_0(\Delta \alpha)
	\right]
\nonumber\\
	&+ \frac{1}{2}\sum_{n \neq 0} \Bigl[
		K_0\bigl(\Delta \sqrt{ x^2 - (vt -i\alpha -i n v \beta \hbar)^2}\bigr)
\nonumber\\
	&\qquad\qquad
		-
		K_0\bigl(\Delta|\alpha+n v\beta\hbar|\bigr)
	\Bigr],
\end{align}
where $\alpha$ is the short regularization length, and where
all square roots are chosen to lie on the first Riemann sheet with positive real parts.
We emphasize that $G^r$ and
$G^a$ do not need regularization as they remain finite when
they approach the light cone. Off the light cone the limit $\alpha \to \eta$ with
infinitesimal $\eta>0$ can be taken too for $W, \tilde{W}$, and $G^t$. In this limit, however,
$G$ would diverge, and it therefore requires a finite regularization $\alpha$
throughout.


\section{Massless case}
\label{sec:massless}

The massless case, $\Delta \to 0$, is a special case of much interest in particular for the study of
one dimensional fermionic conductors described as a Luttinger liquid. Through the method of
bosonization \cite{Haldane1981,vonDelft1998,Gogolin1998,Giamarchi2003,Shankar2017} fermionic correlators
can be expressed in terms of exponents of bosonic correlators, $e^{G(t,x)}$, even in the case of strong
fermionic interactions. This provides a direct access to correlated many-body physics through analytic,
nonperturbative correlation functions, making bosonization a convenient and broadly applied tool,
with numerous applications ranging from quantum transport to correlated many-body dynamics.

Starting from the massive case, the limit $\Delta \to 0$ of $W(t,x)$ is singular though because
$K_0\bigl(\Delta \sqrt{x^2-(vt-i\alpha)^2}\bigr)$ diverges for for all $(t,x)$,
making the finite cutoff $\alpha$ alone insufficient for the regularization. Indeed the newly appearing
divergence is infrared and appears from the singularity of the integrand in Eq.~\eqref{eq:W(t,x)_int}
at $k\to 0$. In bosonization, however, the function $G(t,x) = W(t,x)-W(0,0)$ appears commonly through the identity
$\mean{e^{i\phi(t,x)} e^{-i\phi(0,0)}} = e^{\mean{\phi(t,x)\phi(0,0)}-\mean{\phi^2(0,0)}}=e^{G(t,x)}$
which regularizes naturally the singularity.
This makes $G(t,x)$ the commonly used correlator.

The requirement of an ultraviolet regularization $\alpha$ may actually seem surprising for a condensed matter system
in which the energy bands have an upper limit. The divergence is indeed an artifact from linearizing
the band structure to match fermion-boson identities of correlators \cite{Tomonaga1950,Luttinger1963}, and a
cutoff length $\alpha$ appears naturally through the wavelength of the highest energy states in a band, on the
order of the crystalline lattice spacing. While the low energy physics is universal and well described through
the relativistic theory in 1+1 dimensions, there remains always a non-universal cutoff scale in all expressions,
as an indicator that in one dimension virtual fluctuations to high energy states always have a remnant contribution
even at low energies.
Within standard Luttinger liquid theory this is a lasting feature, but recent developments of nonlinear
Luttinger liquids \cite{Imambekov2012} provide generalizations to overcome this limitation. Such an extension
falls beyond the scope of conventional applications of the Luttinger liquid for which the regularization
as discussed here remains highly relevant.

The coupling to a thermal reservoir introduces through $k_B T$ another energy scale
but of an opposite nature because instead of suppressing fluctuations a large $T$ enhances
incoherent fluctuations to high energies. The dependence on the cutoff thus remains.
But for $k_B T \gg \Delta v \hbar$ we may ask whether the massive and massless theories become
indistinguishable. This question is actually subtle. We will address it in Sec.~\ref{sec:high} and show that the massless
theory provides a good fit at spacetime distances below the thermal length. In Sec.~\ref{sec:GGE} we will then show
that for $\Delta \neq 0$ a massless behavior still appears at zero or low temperatures for small
spacetime distances, not through the massless zero temperature expression but in the form of a transient
generalized Gibbs ensemble with an effective temperature set by the mass term $\Delta$.

Thus the massless theory can be seen as an integral part for describing the massive behavior too.
Since the massless correlators can be given compact expressions at all $T$ without the infinite sums over $n\beta$
we shall provide them here as a benchmark for the transient response in the case of $\Delta \neq 0$.
We therefore rederive in the following results known throughout the literature, but with a special emphasis
on how the analytic properties of the correlators are affected by the cutoff $\alpha$. For the zero
temperature case an analogous careful introduction of $\alpha$ is provided, for instance, in
Ref.~\cite{Shankar2017}. We widen this discussion below to the case of any temperature $T$.

We should just remark that the limit $k_B T > \Delta v \hbar$ may be unphysical for situations in which the
Klein-Gordon form of the action is a harmonic approximation to a more complicated interaction term, such as
for a sine-Gordon theory, in which $\Delta$ itself sets a maximum scale for applicability of the theory.

In the limit $\Delta \to 0$ the eigenmode frequency becomes $\omega_k = v |k|$ and,
from Eq.~\eqref{eq:W(t,x)_int}, we have
\begin{align}
	G(t,x)
	=
	\frac{1}{4}
	\int_{-\infty}^\infty \frac{dk}{|k|}
	\Bigl\{
		 &\bigl[e^{i k x} \cos(t v k)-1\bigr] \coth(\beta \hbar v |k|/2)
\nonumber\\
	&- i e^{i k x}  \sin(t v |k|)
	\Bigr\}.
\end{align}
The subtraction of the $-1$ from the cosine term in the integrand is indeed central to remove the
infrared divergence at $k \to 0$. Further required is the introduction of the ultraviolet regularization
$\alpha$, but as already noted for the massive case, this requires a separate discussion for the
temperature independent and temperature dependent parts.

To perform the splitting into these parts, we use again the expansion
$\coth(z/2) = 1 + 2 \sum_{n\ge 1} e^{-n z}$. The correlator can then be written as
$G(t,x) = G_0(t,x) + 2 \sum_{n \ge 0} G_n(t,x)$, with the temperature independent
term
\begin{align}
	G_0(t,x)
	=
	\frac{1}{4}
	\int_{-\infty}^\infty \frac{dk}{|k|}
	\Bigl\{
		 &\bigl[e^{i k x} \cos(t v k)-1\bigr]
\nonumber\\
	&- i e^{i k x}  \sin(t v |k|)
	\Bigr\} e^{- \alpha |k|},
\label{eq:G_0_massless}
\end{align}
and the temperature dependent terms
\begin{align}
	G_n(t,x)
	=
	\frac{1}{4}
	\int_{-\infty}^\infty \frac{dk}{|k|}
	\bigl[e^{i k x} \cos(t v k)-1\bigr]e^{-n\beta \hbar v |k|}.
\end{align}
Since $G_0$ has an ultraviolet divergence we have introduced in Eq.~\eqref{eq:G_0_massless} the convergence
factor $e^{- \alpha |k|}$, in accordance with Eq.~\eqref{eq:W0_int_regularized} in
the limit $\omega_k/v \to |k|$ for $\Delta \to 0$, as this will preserve the correct analytical structure.
On the other hand, the temperature dependent terms $G_n$ are convergent and, referring to the
discussion at the end of Sec.~\ref{sec:regularization}, the further multiplication of their integrands
by $e^{-\alpha|k|}$ would alter their analytic structure, so that we shall discuss the inclusion of $\alpha$
only after their evaluation.

The temperature independent part can be written as
\begin{equation}
	G_0(t,x) = G_0^+(t,x) + G_0^-(t,x),
\end{equation}
with
\begin{equation}
	G_0^\pm(t,x) = \frac{1}{4}\int_0^{\pm \infty} \frac{dk}{k}
	\bigl( e^{i k (x \mp v t)} - 1 \bigr) e^{-\alpha |k|},
\end{equation}
which expresses modes travelling in the positive ($G_0^+$) and
negative $x$ direction ($G_0^-$), commonly known as right and left movers.
In contrast to the massive system, these are independent from each other.
The solution of the $k$ integrals
is straightforward and leads to
\begin{align}
	\label{eq:G_0_massless_1}
	G_0^\pm(t,x) = \frac{1}{4} \ln\left(\frac{\alpha}{\alpha + ivt \mp i x }\right).
\end{align}
These results allow us to write $G_0 = G_0^+ + G_0^-$ with the same functional dependence on
$\sqrt{x^2-(vt-i\alpha)^2}$ as for the massive case,
\begin{equation}
	\label{eq:G_0_massless_2}
	G_0(t,x) = \frac{1}{2} \ln\left(\frac{\alpha}{\sqrt{x^2-(vt-i\alpha)^2}}\right),
\end{equation}
which matches the asymptotic behavior of Eq.~\eqref{eq:W0_divergent} with both $\Delta^{-1}$ and $\eta$
substituted by $\alpha$,
but it is often recommended to keep the terms $G_0^\pm$ separate for a better control
of the branch cuts of the logarithms as a function of $x \mp v t$.

The temperature corrections can similarly be split into right and left movers
as $G_n = G_n^+ + G_n^-$,
with
\begin{align}
	&G_n^\pm(t,x)
	=
	\frac{1}{8}\int_{-\infty}^{\infty} \frac{dk}{|k|}
	\bigl( e^{i k (x \mp v t)} - 1 \bigr) e^{-n \beta \hbar v |k|}
\nonumber\\
	&=
	\frac{1}{8}\int_{0}^{\infty} \frac{dk}{k}
	\bigl( e^{i k (x \mp v t)} - 1 \bigr) e^{-n \beta \hbar v k}
\nonumber\\
	&+
	\frac{1}{8}\int_{0}^{\infty} \frac{dk}{k}
	\bigl( e^{-i k (x \mp v t)} - 1 \bigr) e^{-n \beta \hbar v k},
\end{align}
which are integrals of the same type as above and evaluate to
\begin{align}
	\label{eq:massless_G_n_plus_minus}
	G_n^\pm(t,x)
	&=
	\frac{1}{8}\ln\left[\frac{(n\beta\hbar v)^2}{(n\beta\hbar v - i x \pm i v t)(n\beta\hbar v + i x \mp i v t)}\right]
\nonumber\\
	&=
	- \frac{1}{8}\ln\left[1 - \frac{(i v t \mp i x )^2}{(n \beta \hbar v)^2} \right].
\end{align}
This leads to
\begin{align}
	2 \sum_{n\ge 1} G_n^\pm(t,x)
	&=
	-\frac{1}{4} \ln\left[ \prod_{n \ge 1} \left( 1 - \frac{(i v t \mp i x )^2}{(n \beta \hbar v)^2} \right) \right]
\nonumber\\
	&
	=
	-\frac{1}{4}
	\ln\left[ \frac{\sin\bigl(\frac{\pi}{\beta\hbar v} (i v t \mp i x ) \bigr)}
	               {\frac{\pi}{\beta\hbar v} (i v t \mp i x )} \right],
\end{align}
where we have used the infinite product formula
$\sin(z) = z \prod_{n\ge 1} (1-z^2/\pi^2n^2)$ \cite[\href{https://dlmf.nist.gov/4.22.E1}{Eq.~(4.22.1)}]{DLMF}.
The term in the denominator of the logarithm cancels the $ivt\mp ix$ contribution of $G_0^\pm$ in
Eq.~\eqref{eq:G_0_massless_1} if we apply now the further shift $vt \to vt - i \alpha$ already discussed
for the massive correlators.
By doing so the full expressions of right and left movers, $G^\pm = G_0^\pm + 2 \sum_{n\ge 1}G_n^\pm$, become
\begin{equation} \label{eq:G_massless_1}
	G^\pm(t,x) =
	\frac{1}{4} \ln\left[\frac{\frac{\pi}{\beta \hbar v} \alpha}{\sin\bigl(\frac{\pi}{\beta \hbar v} (\alpha + i v t \mp i x )\bigr)}\right].
\end{equation}
This result has the correct analytic properties, satisfies the periodicity in the imaginary time direction, and
reproduces the zero temperature limit, Eq.~\eqref{eq:G_0_massless_1}, as $\beta \to \infty$, however, it satisfies $G(0,0) = 0$
only up to order $\alpha$. To produce perfect accuracy we make a further slight adjustment to the regularization in the numerator
inside the logarithm such that
\begin{equation} \label{eq:G_massless_2}
	G^\pm(t,x) =
	\frac{1}{4}
	\ln\left[\frac{\sin\bigl(\frac{\pi}{\beta \hbar v} \alpha\bigr)}{\sin\bigl(\frac{\pi}{\beta \hbar v} (\alpha + i v t \mp i x)\bigr)}\right].
\end{equation}
We shall consider the latter result as the most consistently regularized propagator for left or right movers in the massless case.
This will require though that $2\pi\alpha/\beta \hbar v \ll 1$ to avoid influence of the sine function shape. As this is a natural
requirement of the thermal energy being smaller than the bandwidth it is mostly an appropriate restriction. We should emphasize
though that to order $\alpha$, Eq.~\eqref{eq:G_massless_1}
is as a result as good as Eq.~\eqref{eq:G_massless_2} so that in practice either formula can be employed.

As noted earlier, such a regularization at finite temperature
cannot be obtained by the mere addition of convergence factors $e^{-\alpha |k|}$ in the momentum integrals but is instead
governed by the compatibility with analytic structure and limiting values. We emphasize
though that the differences between such regularizations are on the order of $\alpha^2$ such that for practical purposes
the regularization by $e^{-\alpha |k|}$ or by any other method does not change the accuracy of the results, unless the dependence
of the precise analytic structure of the correlators in the complex plane is essential.

The sum of right and left mover contributions finally leads to
\begin{align}
	&G(t,x)
	=
	G^+(t,x) + G^-(t,x)
\nonumber\\
	&=
	\frac{1}{4}\ln\left[
		\frac{\sin\left(\frac{\pi\alpha}{\beta\hbar v}\right)}%
		     {\sin\left(\frac{\pi(\alpha+ivt - i x)}{\beta\hbar v}\right)}
		\frac{\sin\left(\frac{\pi\alpha}{\beta\hbar v}\right)}%
		     {\sin\left(\frac{\pi(\alpha+ivt + i x)}{\beta\hbar v}\right)}
	\right].
\label{eq:complete_massless_G}
\end{align}
This result is valid for all $(t,x)$.


\section{High temperature limit}
\label{sec:high}

For a massive system, it is natural to question whether at high temperatures, where $k_B T \gg \hbar v \Delta$, the thermal
fluctuations completely overrule the effect of the mass, such that the resulting correlators are well approximated by the massless
results derived in Sec.~\ref{sec:massless}. We shall demonstrate in this section that this only holds at short times and
distances whereas otherwise the mass term $\Delta$ has a persistent influence.

To demonstrate this, we notice that in the high-temperature limit, $\beta \to 0$, it follows from Eqs.~\eqref{eq:fluct_diss} and
\eqref{eq:fluct_diss_1} that as long as $\omega$ is such that $\hbar |\omega| /k_B T \ll 1$ we have
\begin{align}
	W(\omega,k) &\sim 2\pi\frac{k_{B}T}{\omega} A(\omega,k),
\label{eq:fluct_diss_high_T}
\\
	\tilde{W}(\omega,k) &\sim 2\pi\frac{k_{B}T}{\omega} A(\omega,k),
\label{eq:fluct_diss_high_T1}
\end{align}
whose sum, through Eq.~\eqref{eq:fluct_diss_2}, produces thus the classical fluctuation-dissipation dependence on $T/\omega$.
In this limit the Fourier transform of both $k \to x$ and $\omega \to t$ yields
\begin{equation} \label{eq:W_high_T_in_A}
	W(t,x) \sim k_{B}T\int_{-\infty}^{\infty}\frac{d\omega}{\omega} \ e^{-i\omega t} A(\omega, x).
\end{equation}
This equation is valid if the integral is dominated by frequencies such that
$\hbar|\omega| \ll k_B T$. As $A(\omega,x)$, through the integration over $\omega_k$, is nonzero
only for $|\omega| > v \Delta$, the high temperature limit thus needs to fulfil two conditions,
first that $\hbar v \Delta \ll k_B T$ and second that $t \gg \hbar/k_B T$, as $t$ is conjugate
to $\omega$.

To evaluate the $\omega$ integration under these conditions, we further remark that since
$v \Delta$ sets the characteristic frequency of the integrand we can crudely estimate the
integral as
\begin{align}
	W(t,x)
	&\sim \frac{k_{B}T}{v \Delta}\int_{-\infty}^{\infty}d\omega\ e^{-i\omega t} A(\omega, x)
\nonumber\\
	&= \frac{k_B T}{\hbar v \Delta} \bigl[ W_0(t,x) + W_0(-t,x) \bigr].
\end{align}
A different way of seeing this is to notice that
$W_n(t,x) \approx W_0(t,x)$ whenever the imaginary time shift is small, which means
if the integer $n$ is such that
$n v \Delta \beta \hbar < 1$. In this limit $W_n(t,x) \approx [W_0(t,x) + W_0(-t,x)]$,
where the two terms arise from the limits
$\xi_n^\pm = \Delta \sqrt{ x^2 - v^2(t\pm i n \beta \hbar)^2} \approx \xi_n^\pm
= \Delta \sqrt{ x^2 - v^2(t\pm i \alpha)^2} = \Delta \sqrt{ x^2 - v^2(\mp t- i \alpha)^2}$,
keeping the $i \alpha$ to avoid issues with the analytic continuity.
As this approximation holds up to $n \sim 1/v \Delta \beta \hbar = k_B T / \hbar v \Delta$
and for larger $n$ the terms are suppressed exponentially we find again the estimate
$W(t,x) = W_0(t,x) + \sum_{n \ge 1} W_n(t,x) \sim \frac{k_B T}{\hbar v \Delta} \bigl[ W_0(t,x) + W_0(-t,x) \bigr]$.

These considerations indicate that the high temperature limit can be obtained from the
zero temperature expressions in a rather straightforward way. To improve on how $(t,x)$
further affect the dependence on $k_B T/\hbar v \Delta$ we can use, for instance,
$1/\omega = \int_0^\infty d\tau \, e^{-\tau \omega}$, and write
Eq.~\eqref{eq:W_high_T_in_A} as
\begin{equation}
	W(t,x) \sim k_{B}T \int_0^\infty d\tau \int_{-\infty}^{\infty} d\omega \ e^{-i\omega (t-i\tau)} A(\omega, x),
\end{equation}
or more carefully
\begin{align}
	&W(t,x) \sim \,
	k_{B}T\lim_{\eta \to 0^{+}}\int_{\eta}^{\infty}d\tau \int_{0}^{\infty}d\omega\ e^{-i\omega (t-i\tau)}A(\omega, x)
\nonumber\\
  &-\,  k_{B}T\lim_{\eta \to 0^{+}}\int_{\eta}^{\infty}d\tau\int_{-\infty}^{0}d\omega\ e^{-i\omega (t+i\tau)}A(\omega, x),
\end{align}
where $\eta$ acts as before to preserve causality.
If we assume that $A(\omega,k) = -A(-\omega,k)$, as consistent with Eq.~\eqref{eq:A(w,k)_gapped}, then the substitution
$\omega \to -\omega$ in the second integral and making use of Eqs. \eqref{eq:fluct_diss_limit} and \eqref{eq:fluct_diss_limit_1} leads to
\begin{equation} \label{eq:W_high_T}
	W(t,x) \sim \frac{k_{B}T}{\hbar}\int_{0}^{\infty}d\tau\bigl[W_{0}(t-i\tau,x) + W_{0}(-t-i\tau,x) \bigr],
\end{equation}
where the $\eta \to 0^+$ limit is captured by the cutoff $\alpha$ in $W_0$.

\begin{figure}
	\centering
	\includegraphics[width=\columnwidth]{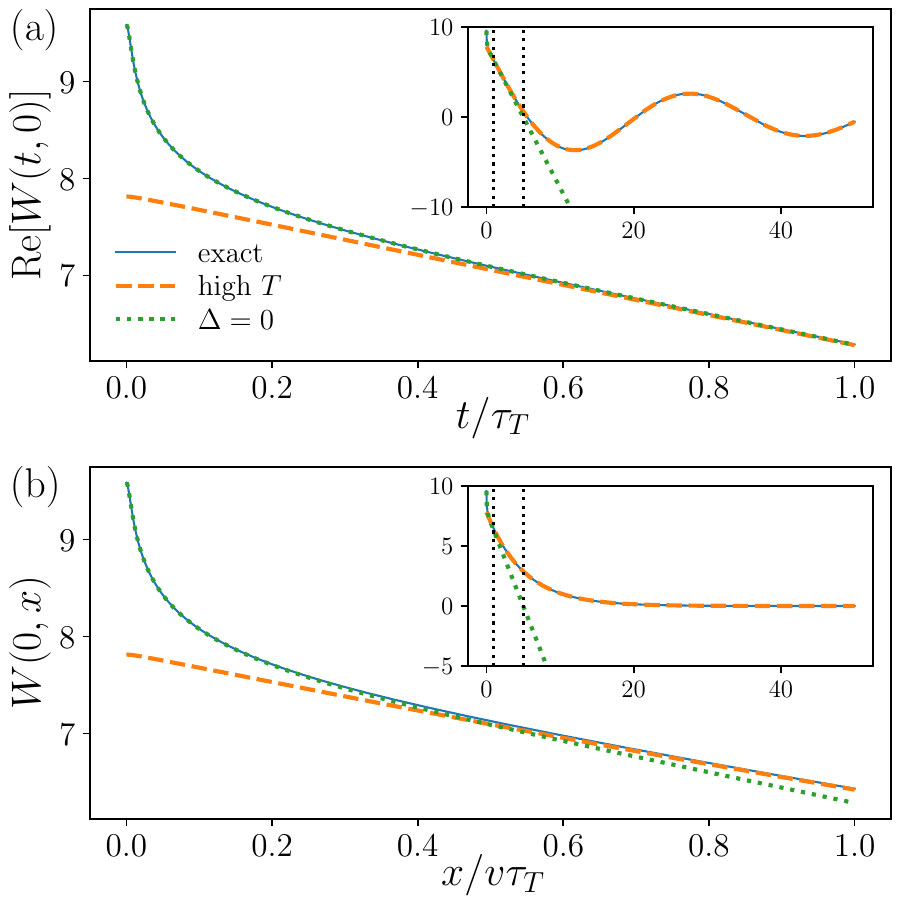}
	\caption{\label{fig:W_high_T}%
		Comparison between the exact Wightman function $W(t,x)$ of Eq.~\eqref{eq:W_full} (solid blue lines),
		the high temperature ($k_B T \gg \hbar v \Delta$) approximation of Eq.~\eqref{eq:W_high_T} (dashed orange lines),
		and the massless solution with $\Delta=0$ obtained from Eq.~\eqref{eq:complete_massless_G} shifted by a constant
		such that it matches $W(0,0)$ at the origin.
		Chosen are $k_B T = 5\, \hbar v \Delta$ and $\alpha = 10^{-3}/\Delta$, with $\Delta$ setting the unit of inverse length.
		The $n$ summation in Eq.~\eqref{eq:W_full} is terminated at $n = 10\, k_B T /\hbar v \Delta = 50$.
		Panel (a) shows the behavior as a function of $t/\tau_T$ at $x=0$ with $\tau_T = \hbar / k_B T$ the thermal time,
		and panel (b) as a function of $x/v\tau_T$ at $t=0$.
		The curves show that the high temperature approximation indeed matches the exact function at $t, x/v \gtrsim \tau_T$, with the
		insets showing the perfect agreement at larger $t$ and $x$ values, respectively.
		The massless approximation, however, matches the exact function at $t, x/v \lesssim \tau_T$ instead.
		In panel (a) the vertical dotted lines mark the characteristic times $t = \tau_T$ and $t=1/v \Delta = 5\, \tau_T$,
		and in panel (b) the corresponding characteristic distances $x = v \tau_T$ and $x = 1/\Delta = 5\, v \tau_T$.
	}
\end{figure}

Figure~\ref{fig:W_high_T} shows for $k_B T = 5 \, \hbar v \Delta$
comparisons between the exact $W(t,x)$, shown as solid blue lines and
obtained by Eq.~\eqref{eq:W_full} with the summation
carried out up to $n = 10 k_B T / \hbar v \Delta$ to ensure convergence,
and the high temperature asymptote of Eq.~\eqref{eq:W_high_T},
shown as dashed orange lines.
In panel (a) we plot $\Real[W(t,x)]$ as a function of $t$ at $x=0$ and confirm that indeed the high temperature approximation
becomes highly accurate
at $t \gtrsim \tau_T = \hbar / k_B T$.
A similar behavior would be shown by $\Imag[W(t,0)]$.
Panel (b) shows the same comparison at $t=0$ as a function of $x$ instead (noting that $W(0,x)$ is real).
Although the approximation has been justified only by considering large enough values of $t$,
it is notable that even at $t=0$ the condition $x \gtrsim v\tau_T$ leads to a similar accuracy.
This is indeed because the approximation $\omega \sim \omega_k \sim v\Delta$ focuses on small $k$, such that
a good matching for large $x$ can be expected.
Instead of diverging to $-\infty$ as the massless result of Eq.~\eqref{eq:complete_massless_G} suggests, the mass term forces
the correlation functions to decay to zero, irrespective of the temperature.

Temperature dominates over $\Delta$ instead at short $t,x/v < \tau_T$. This is visualized in Fig.~\ref{fig:W_high_T} by the green dotted lines
which shows $W(t,x)$ at $\Delta=0$ for the same temperature, obtained by Eq.~\eqref{eq:complete_massless_G} and the addition of
constant $W(0,0)$
to match the initial value of the exact $W(t,x)$. The correspondence for $W(t,0)$ is excellent up to $t \sim \tau_T$ but then splits off,
whereas for $W(0,x)$ the accuracy extends to $x \sim v \tau_T/2$. At very small $(t,x)$ all expressions show the temperature independent
logarithmic behavior, but over the range up to $\tau_T$ it is indeed the full temperature dependent massless expression that provides
the most accurate approximation.
This has an intriguing underlying physical reason that shall be investigated next.


\section{Transient generalized Gibbs ensembles}
\label{sec:GGE}

We now consider the transient behavior of $W(t,x)$ around the spacetime origin, starting with the limit of low temperatures,
where $k_B T \ll \hbar v \Delta$. In this case, the summation over $n$ in Eq.~\eqref{eq:W_full} can either be restricted to
a few terms or neglected entirely due to the exponential suppression that arises from the asymptotics
of the modified Bessel function, $K_0(z) \sim \sqrt{\pi/(2z)} e^{-z}$ for $|z|\to \infty$
\cite[Eq.~\href{https://dlmf.nist.gov/10.25.E3}{(10.25.E)}]{DLMF}. For practical purposes it is thus sufficient to neglect
temperature effects throughout and consider $W(t,x) \approx W_0(t,x)$.

While at large spacetime separations the function $W_0(t,x)$ is well described by the asymptotics of the Bessel function,
which, as was shown in Fig.~\ref{fig:W0}, leads to an algebraic decay with oscillations for time-like arguments and an exponential
decay for space-like arguments, the behavior at short spacetime separations is significantly more intriguing.

\begin{figure}
	\centering
	\includegraphics[width=\columnwidth]{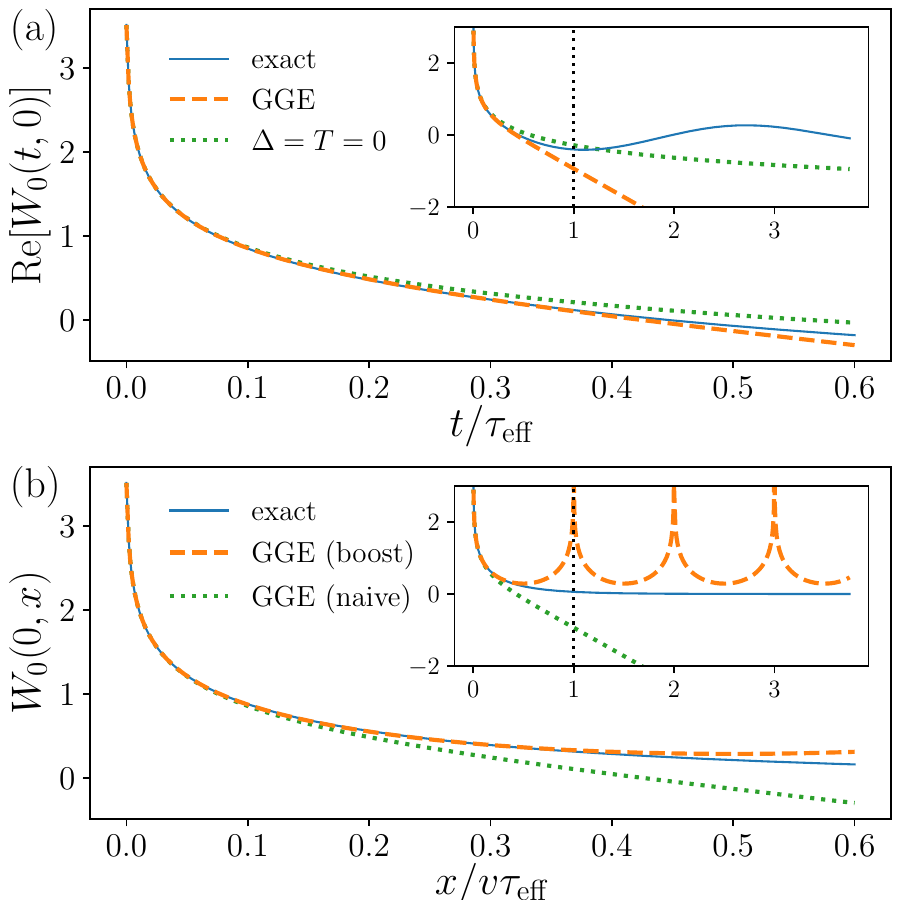}
	\caption{\label{fig:W_GGE}%
	Panel (a) shows the exact solution of $\Real[W_0(t,x=0)]$ at $T=0$, $\Delta\neq 0$ (solid blue line),
	the GGE approximation of Eq.~\eqref{eq:W_GGE_t}
	consisting of the massless theory with effective temperature $T_\text{eff} = \hbar v \Delta / 2 k_B$
	(dashed orange line), and the naive approximation of the zero temperature, zero mass solution
	obtained from Eq.~\eqref{eq:G_0_massless_2}, shifted by the exact $W(0,0)$ (dash dotted green line).
	The time axis is scaled to the characteristic time $\tau_{\text{eff}} = \hbar / k_B T_\text{eff} = 2/v\Delta$.
	While all curves show the same logarithmic behavior at very short $t$, at intermediate times reaching
	up to $\sim 0.4 \tau_\text{eff}$ the GGE curve provides a more accurate
	approximation than the simple massless limit, as corroborated by the arguments in the main text.
	The inset shows the extension to longer times where the approximate solutions clearly fail
	beyond the transient regime set by $\tau_\text{eff}$ (marked by the vertical dotted line).
	Panel (b) compares the exact, real solution in the space-like direction $W(t=0,x)$ with the axis
	scaled to the effective length $v \tau_\text{eff}=2/\Delta$.
	Shown is the exact solution (solid blue line),
	the GGE approximation of Eq.~\eqref{eq:W0_GGE_general} obtained through a Lorentz boost from the $t$ dependence (dashed orange line),
	and the naive GGE (dotted green line) obtained by the standard $(t,x)$ dependence of the
	massless theory, Eq.~\eqref{eq:complete_massless_G},
	at effective temperature $T_\text{eff}$. The latter does not maintain Lorentz
	invariance and clearly loses accuracy much earlier than the boosted GGE.
	The $x$ axis is scaled to the characteristic length $v \tau_\text{eff} = 2/\Delta$.
	The inset shows the extension to larger $x$ values, with length $v \tau_\text{eff}$ marked
	by the vertical dotted line.
	}
\end{figure}

Let us first focus on the behavior of $W_{0}$ at $x=0$ as we vary $t$. We observe through Figs.~\ref{fig:W0}
and \ref{fig:W_high_T} similarity in the structures of the zero-temperature massive and finite-temperature
massless correlation functions at short times. Although at small arguments the Bessel function is logarithmic,
$K_0(z) \sim -\ln(z)$, which resembles the zero temperature limit of the massless case,
Eq.~\eqref{eq:G_0_massless_2}, we find that it is in fact much better approximated by a massless theory at
a finite effective temperature given by $T_\text{eff} = \hbar v \Delta / 2k_B $. Indeed, we show in
Fig.~\ref{fig:W_GGE} (a) that by plugging $T_\text{eff}$ into Eq.~\eqref{eq:complete_massless_G} we recover
an excellent approximation to $W_{0}(t,0)$ via
\begin{equation}
	W_{0}(t,0)
	\approx
	W_{0}(0,0)
	+
	\frac{1}{4}\ln\left[
		\frac{\sin^2\left(\frac{\pi\alpha}{\beta_\text{eff}\hbar v}\right)}%
		     {\sin^2\left(\frac{\pi(\alpha+ivt)}{\beta_\text{eff}\hbar v}\right)}
	\right],
\label{eq:W_GGE_t}
\end{equation}
where $\beta_\text{eff} = 1/k_B T_\text{eff} = 2/\hbar v \Delta$, $W_{0}(0,0) = \frac{1}{2} K_0(\Delta \alpha)$,
and the range of agreement is
set by the characteristic timescale $\tau_{\text{eff}} = \hbar\beta_\text{eff} = \hbar/ k_B T_\text{eff}$.

While a massless behavior of $W(t,0)$ may be expected due to strong quantum fluctuations at short times, the
effective temperature takes its origin in the fact that the fields $\phi(t,x)$ create excitations out of the
ground state of the massive system. In contrast to a massless theory, the mass term binds the right and
left moving modes together, causing them to be maximally entangled.
Consequently the excitations monitored by $W_0(t,0)$ occur simultaneously for right and left movers, such that
left movers act immediately as a bath on the right movers, and vice-versa. By the principle of maximum entropy \cite{Jaynes1957}
such a bath mimics a thermal bath, but with a temperature $T_\text{eff}$ set by the characteristic energy scale
of the correlated system, $\Delta$. Hence it is possible to describe the \emph{transient} behavior by a generalized Gibbs ensemble (GGE),
in which expectation values are drawn from a Gibbs ensemble that is determined by the artificial temperature $T_\text{eff}$.
This is in stark contrast to the usual occurrence of the GGE, which has been introduced to describe \emph{steady} states
after a quench in integrable systems in which conserved quantities prevent a full
thermalization \cite{Rigol2007,Caux2012,Langen2015,Vidmar2016,Jaynes1957}, yet recent developments have
pointed out that the GGE also can appear through subsystem entanglement in gapped ground states \cite{Moghaddam2022}.

To elaborate this physics in detail we first need to specify the origin of the mass term $\Delta$. As noted in
Sec.~\ref{sec:massless}, the massive theory is not smoothly connected to the massless case, but is instead
qualitatively different in that the right and left movers are
bound to each other. To achieve this coupling, the underlying Hamiltonian must have a large-energy component
through which the mass emerges. For an elementary field theory, this can result for instance from a Higgs mechanism. For an
effective many-body theory this can arise from a sine-Gordon interaction term that  pins $\phi$ to its minimum,
such that harmonic fluctuations that arise from perturbations from the kinetic contributions to the Hamiltonian
are captured by an effective mass term. With this understanding, we can write the Hamiltonian in the form
$H = H_\Delta + H_+ + H_-$, where $H_\Delta$ sets the large energy scale, and $H_{\pm}$ are the kinetic perturbations
arising for the right ($+$) and left ($-$) movers respectively. We emphasize that $H_\Delta$ can not just contain
the quadratic mass term $\propto \Delta^2 \phi^2$, as without kinetic terms
there does not exist a stable ground state. Instead, $H_\Delta$ must be able to localize $\phi$ to a distinct
value and create an energy gap between the ground and the excited states.

Let us consider, for instance, the sine-Gordon model with the action
\begin{align}
	S_{SG}
	&= \frac{\hbar}{2\pi v} \int dt\, dx\, \phi(t,x)
	\bigl( -\partial_t^2 + v^2 \partial_x^2  \bigr) \phi(t,x)
\nonumber\\
	&
	+
	\frac{\hbar v}{4\pi} \Delta^2
	\int dt\, dx\, \cos\bigl( 2\phi(t,x)\bigr).
\label{eq:SG}
\end{align}
which leads to the considered Klein-Gordon model, Eq.~\eqref{eq:S}, when expanding the cosine term
to quadratic order in $\phi$ about its minimum. As shown in Refs.~\cite{Luther1974,Coleman1975}
this model can be mapped onto a Dirac fermion model with spectrum $\pm \hbar v \sqrt{k^2+\Delta^2}$ exhibiting a
gap of $2 \hbar v \Delta$ when $k=0$. Without the kinetic
contribution, the spectrum is $k$ independent and forms flat bands at energies $\pm \hbar v \Delta$.

In typical GGE setups, a quench is produced from an equilibrium state of $H_\Delta + H_+ + H_-$ by suddenly switching
off the $H_\Delta$ term at time $t=0$, such that further evolution is under $H_+ + H_-$ only \cite{Rigol2007,Caux2012,Vidmar2016}.
Since, following the quench,
the right and left moving modes are no longer coupled, they tend to a steady state that, for a system as considered here,
is described by a product of density matrices $\rho_\pm$ corresponding to the right and left movers respectively.
Each density matrix is then given in terms of a GGE as
\begin{equation}
	\rho_\pm = e^{- H_\pm / k_B T_\text{eff}}/Z_\pm,
\end{equation}
where the partition function is given by $Z_\pm = \Tr_\pm\{ e^{- H_\pm / k_B T_\text{eff}}\}$
and $\Tr_\pm$ represents the trace over the right and left moving degrees of freedom respectively.

Although this scenario does not match our situation, as we do not switch off $H_\Delta$, we still observe in
Fig.~\ref{fig:W_GGE} (a) a time evolution of $W(t,0)$ that is effectively massless and, for a transient regime
surrounding the spacetime origin, well described by the GGE states $\rho_\pm$ of decoupled right and left movers.
This reveals two notable features of the GGE. First, that it occurs already in an excitation out of the ground
state as produced by $\phi(0,0)$, instead of requiring the full quench with setting $H_\Delta=0$. Second, it
reveals that the GGE physics sets in almost immediately and not only in the long time limit. Indeed $H_\Delta$
is now not only responsible for defining the initial state, but also for forcing the system to relax to that
same state at long times. Therefore, the GGE has to be invalidated at a time scale set by $\tau_\text{eff}$,
which is indeed what we observe in Fig.~\ref{fig:W_GGE}. It is thus remarkable that the GGE description is
already fully developed at short times $t \ll \tau_\text{eff}$, such that the GGE with its effective temperature
$T_\text{eff}$ is fully appreciable in this transient regime, without the need for drastic quenches that modify the Hamiltonian.

To obtain quantitatively the temperature $T_\text{eff}$ and the states $\rho_\pm$ we
follow Refs.~\cite{Peschel2011,Qi2012}. The advantage for the transient regime is that the identification
with fully developed density matrices $\rho_\pm$ does not need to be perfect as we are interested only
in $t /\tau_\text{eff} < 1$. We thus proceed with perturbation theory, by considering $H_\pm/k_B T_\text{eff}$ as a small parameter
such that $e^{-H_\pm/k_B T_\text{eff}} \approx \mathbbm{1}_\pm - H_\pm/k_B T_\text{eff}$, where $\mathbbm{1}_\pm$
is the identity on the $\pm$ states. Furthermore, we have $Z_\pm \approx N_\pm (1 - \mean{H_\pm}_0/k_B T_\text{eff})$,
where $N_\pm = \Tr_\pm\{\mathbbm{1}_\pm\}$ is
the number of $\pm$ states and $\mean{H_\pm}_0 = \Tr_\pm\{ H_\pm\}/N_\pm$ is the expectation value over the unperturbed
ground state, as we shall see.
Collecting these terms for $\rho_\pm$ to first order in $H_\pm$, our aim is thus to show that
\begin{equation} \label{eq:rho_pm_expansion}
	\rho_\pm \approx \frac{1}{N_\pm} \left( \mathbbm{1}_\pm - \frac{H_\pm - \mean{H_\pm}_0}{k_B T_\text{eff}}\right).
\end{equation}
We will outline here the general approach \cite{Peschel2011} and provide a concrete calculation
in \ref{sec:massive_Dirac}. We shall construct the density matrices from the
ground state wave function to first order in perturbation in $H_\pm$.
Since the infinite system is diagonal in momentum $k$ we can restrict our attention
to a subspace of fixed $k$. Within such a subspace let
$\ket{\Phi_n}$ be an eigenbasis of the unperturbed $H_\Delta$ with $\ket{\Phi_0}$ the ground state.
The perturbed ground state $\ket{\Psi_0}$ is then obtained to first order in $H_\pm$ by
\begin{equation}
	\ket{\Psi_0}
	=
	\ket{\Phi_0}
	-
	\sum_{n \neq 0}
	\ket{\Phi_n}\frac{\bra{\Phi_n} \bigl( H_+ + H_- \bigr) \ket{\Phi_0}}{E_n-E_0},
\end{equation}
where $E_n$ are the eigenenergies of $H_\Delta$. Under the assumption that the system
has a single excitation energy we can replace the energy difference by
$E_n-E_0=2 \hbar v \Delta$. Since the system has right/left symmetry and
we have equality of the averages
$\bra{\Phi_n} H_+ \ket{\Phi_0} = \bra{\Phi_n} H_- \ket{\Phi_0}$
for $n \neq 0$ as a characteristic of their maximal entanglement.
If we use furthermore
$\sum_{n \neq 0} \ket{\Phi_n}\bra{\Phi_n} = \mathbbm{1} - \ket{\Phi_0}\bra{\Phi_0}$,
with $\mathbbm{1}$ the identity, we
can eliminate the sum over the excited states $n$ and
obtain
\begin{equation}
\label{eq:Psi_0}
	\ket{\Psi_0}
	= \ket{\Phi_0} - \frac{1}{\hbar v \Delta} \tilde{H}_\pm \ket{\Phi_0},
\end{equation}
where we have set $\tilde{H}_\pm = H_\pm - \bra{\Phi_0} H_\pm \ket{\Phi_0} \mathbbm{1}_\pm$
and used the equality $\tilde{H}_+ \ket{\Phi_0}=\tilde{H}_- \ket{\Phi_0}$
which follows from the maximal entanglement.

With this result the density matrix for the ground state, $\rho$, becomes to first order
\begin{align}
	\rho
	&= \ket{\Psi_0}\bra{\Psi_0}
	=
	\ket{\Phi_0}\bra{\Phi_0}
\nonumber\\
	&	- \frac{1}{\hbar v \Delta}
	\Bigl[
		\ket{\Phi_0}\bra{\Phi_0}
		\tilde{H}_\pm
		+
		\tilde{H}_\pm
		\ket{\Phi_0}\bra{\Phi_0}
	\Bigr].
\end{align}
To make the link with the GGE we consider the bipartition of the system
between the right and left movers. By tracing out the degrees of freedom for the left movers,
we obtain a reduced density matrix for the right movers only,
$\rho_+ =\Tr_-\{\rho\}$.
The initial reduced density matrix is $\rho_+^{0}= \Tr_-\{\ket{\Phi_0}\bra{\Phi_0}\}$, and
the property of maximal entanglement means that all states are equally populated such that
$\rho_+^0 = \mathbbm{1}_+/N_+$. As in addition $\tilde{H}_+$ does not depend on the $-$ modes
it can be taken out of the trace, and we obtain
\begin{align}
	\rho_+
	=
	\Tr_-\{\rho\}
	=
	\frac{1}{N_+}
	\left(
	\mathbbm{1}_+
	- \frac{2\tilde{H}_+}{\hbar v \Delta}
	\right).
\end{align}
Finally we note that $\bra{\Phi_0} H_+ \ket{\Phi_0} = \Tr\{H_+ \ket{\Phi_0}\bra{\Phi_0}\} = \Tr_+\{H_+\}/N_+ = \mean{H_+}_0$,
which produces a match with Eq.~\eqref{eq:rho_pm_expansion} upon the identification
$k_B T_\text{eff} = \hbar v \Delta / 2$.
The matching for $\rho_-$ is obtained in the same way by changing the $+$ indices to $-$.
This confirms the use of the GGE as described above.

As the GGE arises as a ground state property it remains valid for the propagation until the perturbation $H_+ + H_-$
has had enough time to invalidate the perturbative ansatz. This explains why the GGE description sets in abruptly
after the excitation but breaks down after the characteristic time scale $\tau_\text{eff}$.

\begin{figure}
	\centering
	\includegraphics[width=\columnwidth]{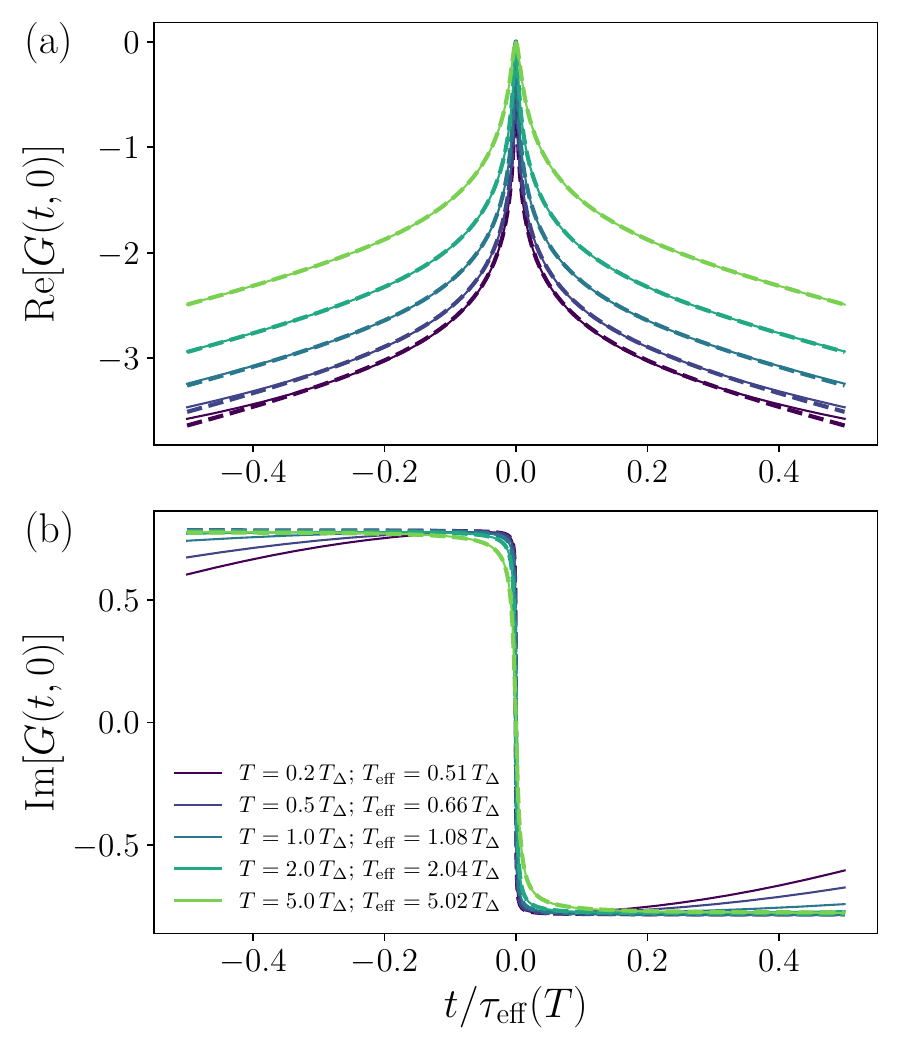}
	\caption{\label{fig:GGE_T}%
		Display of the exact $G(t,0) = W(t,0)-W(0,0)$ (solid lines) at various temperatures (colors changing from
		dark to bright with increasing temperature) together with the GGE approximation (dashed lines of same color)
		of zero mass and effective temperature $T_\text{eff}(T)$ as given by Eq.~\eqref{eq:T_eff}.
		The cutoff is chosen as $\alpha = 10^{-3}/\Delta$.
		Panel (a) shows the real and panel (b) the imaginary parts of the functions. The time axis is scaled
		by the characteristic time $\tau_\text{eff}(T) = \hbar /k_B T_\text{eff}(T)$ for each curve. Real and effective temperatures
		are as indicated in panel (b) in comparison with the temperature $T_\Delta = \hbar v \Delta/k_B$ set by
		the mass term. The agreement between exact and GGE at $t \ll \tau_\text{eff}(T)$ is remarkable,
		and extends to sizable portions of $\tau_\text{eff}(T)$ that increase with $T$.
	}
\end{figure}

For the massive Dirac fermion model, or equivalently the sine-Gordon model, it is possible to obtain the exact
stationary GGE expressions following the quench $H_\Delta \to 0$ at any physical temperature.
This yields an effective temperature that depends on both $\Delta$ and the real temperature $T$, given by \cite{Chung2012}
\begin{equation} \label{eq:T_eff}
	k_B T_\text{eff}(T) = \frac{\hbar v \Delta / 2}{\tanh(\hbar v \Delta / 2 k_B T)}.
\end{equation}
Notably, $T_\text{eff}(T)$ interpolates between $T_\text{eff}(0) = \hbar v\Delta/2$, as identified earlier, and
$T_\text{eff}(T) \approx T$ at large temperatures. Although this result has been derived for the stationary regime
long after the quench \cite{Chung2012}, and with the excitation by $\phi$ we do not produce an equivalent quench,
Fig.~\ref{fig:GGE_T} shows that the corresponding GGE expressions remain valid at any temperature  $T$ for a short
period of time set by the now temperature-dependent characteristic scale $\tau_\text{eff}(T) = \hbar / k_B T_\text{eff}(T)$.
Furthermore these GGE expressions are seen to extend over longer
ranges with increasing $T$. With the $T_\text{eff}(T) \approx T$ large temperature limit, this also covers the high
temperature case discussed in Sec.~\ref{sec:high}.
The remarkable matching to the GGE description with effective temperatures $T_\text{eff}(T)$ for
the full range of real temperatures $T$ shows that the appearance of the transient GGE is
indeed a robust physical feature and not a coincidence at zero temperature.

In our description of the transient GGE we have so far only taken the $t$ dependence into account and set $x=0$.
We now question how an $x$ dependence can be fed into this picture. When $k_B T \ll \hbar v \Delta$ we have
already noted that the sum over $n$ in Eq.~\eqref{eq:W_full} can be neglected due to the exponential suppression
that feeds in through the Bessel function. In this case, what remains is a function $W_{0}$ that is, away from
the light cone, Lorentz invariant. We therefore expect the transient GGE to also inherit this Lorentz invariance
in the low-temperature limit.
This means that instead of using Eq.~\eqref{eq:complete_massless_G} we perform a Lorentz boost
$-t + i \alpha \to \sqrt{x^2 - (t-i\alpha)^2}$ in Eq.~\eqref{eq:W_GGE_t} such that, at $k_B T \ll \hbar v \Delta$,
\begin{equation} \label{eq:W0_GGE_general}
	W_0(t,x) \approx W_0(0,0) + \frac{1}{4}
	\ln\left[\frac{\sin^2\left(\frac{\pi\alpha}{\beta_\text{eff}\hbar v}\right)}
	              {\sin^2\left(\frac{\pi\sqrt{x^2-(vt-i\alpha)^2}}{\beta_\text{eff}\hbar v}\right)}\right],
\end{equation}
where $\beta_\text{eff} = 1/k_B T_\text{eff}(0)$. We demonstrate the agreement of this approximation to the precise
form for $W_0$ at $t=0$ as $x$ is varied in Fig.~\ref{fig:W_GGE} (b), for which we find once more
agreement up to a distance scale set by $v\tau_\text{eff}(0)$.

The argument of Lorentz invariance can no longer be used at temperatures $k_B T \sim \hbar v \Delta$ or higher.
Although the discussion of high temperatures in Sec.~\ref{sec:high} has shown that a massless transient behavior remains
valid, this no longer follows the boost of Eq.~\eqref{eq:W0_GGE_general} but has an $x$ dependence that matches better
the genuinely massless Eq.~\eqref{eq:complete_massless_G}.
Therefore, we should not expect a clear GGE behavior to extend to this physical situation.
It certainly would be possible to devise an interpolation between the limiting cases, but there would
no longer be the clear physical picture we could employ for the $t$ dependence. Thus such an interpolation
would be rather formalistic, and we prefer to not pursue it.

To close this section, let us return to the original motivating example, Eq.~\eqref{eq:W_GGE_t}. One qualitative
change following the introduction of the mass-dependent effective temperature that needs to be addressed,
particularly in this low-temperature limit, is the periodicity of the approximation under imaginary time shifts
$t \to t - i n \beta_\text{eff}(0) \hbar$ for $n \in \mathbb{Z}$, giving rise in frequency space to a dependence
on bosonic Matsubara frequencies of the form
$\omega_n = 2 \pi n k_B T_\text{eff}(0)/\hbar$. As $T_\text{eff}(0) = \hbar v \Delta/2$, we deem it appropriate
to label these $\omega_n$ as \emph{massubara frequencies}, and since the GGE expression is valid only at short
times, a large number of massubara frequencies will be required for its reproduction.

This contrasts the exact $W(t,x)$ though, which has a substantially different analytic structure in frequency space.
The finite-temperature Matsubara periodicity is different, and at $T=0$ in particular it is absent.
Furthermore, from the dependence of the spectral function $A(\omega,k)$ on $\delta(\omega \pm v \sqrt{k^2+\Delta^2})$
it follows that $W(\omega,x)$ depends on $\sqrt{\omega^2-v^2\Delta^2}$, which has a branch cut, together
with the non-analytic condition $\omega > \Delta$. The analysis of such a structure in complex $\omega$ space is
challenging, and it is thus remarkable that through the GGE's massubara frequencies there exists an approximation
for a range of $(t,x)$ values that produces an excellent match while being conventional and more readily treatable
in its analytic structure. For practical calculations we therefore expect a broader practical applicability of the GGE behavior
beyond the fundamental observations exposed earlier.


\section{Profile function approximation}
\label{sec:profile}

In the previous two sections we demonstrated that there are excellent approximations to $W(t,x)$ for the long or short
spacetime behaviors at high and low temperatures. While these were obtained by analyzing the physical properties of
the theory, they can as well be considered as mathematical approximations for Bessel functions that go beyond standard
series expansions. It is notable though that there is a qualitative difference between the small and large argument limits
that could not be captured by the methods introduced above. Indeed the
long spacetime (high temperature) approximation of Sec.~\ref{sec:high} neglects the quantum fluctuations at short scales,
whereas the transient GGE introduced in Sec.~\ref{sec:GGE} breaks down when the pinning by the mass term undoes the
stirring up caused by the quench-like excitation by $\phi(0,0)$. The cross-over between the clear, simple physical
processes that describe the asymptotics requires then the description of a special function, which is precisely provided
by the Bessel function $K_0$.

Such an exact expression, however, may become cumbersome to use for further treatment
where the two-point correlators are just intermediate steps in a longer evaluation,
such as in diagrammatic expansions
if the frequency-momentum representation is unsuitable, or if further integration over powers and products of $e^{G(t,x)}$
are required as often found in bosonization calculations.
Although with the exact functions specific solutions can always be computed to arbitrary numerical accuracy,
this does not reveal anything about functional dependences on system parameters and thus may not allow to gain direct access
to the involved physics.
If the previously derived approximations are too restrictive,
then it is desirable to find yet another approximation that extends over all $(t,x)$ but is composed of elementary functions.
Indeed, if integration over $x$ or $t$ is required then the limits of large or small arguments are insufficient and
an approximation spanning over all space-time becomes necessary.
The most important feature we want to carry over from the precise results given in Sec.~\ref{sec:summary_analytic} is
the non-perturbative dependence on $\Delta$, since indeed the opening of a gap changes the topology
of the dispersion relation. As we have seen, the massless and massive theories are fundamentally different such that,
for instance, a perturbative expansion in $\Delta$ could never correctly capture the physics. The price we will have to pay,
however, is to sacrifice the GGE behavior
discussed in Sec.~\ref{sec:GGE}. Since the GGE expressions are described by a massless theory they would always fail
in the limit of large arguments, as shown in Fig.~\ref{fig:W_GGE}.

We will henceforth focus on the zero temperature, $T=0$, case and outline later the reason why
implementing temperature corrections into the proposed scheme would be tricky. It follows that
$W(t,x) = W_0(t,x) = \frac{1}{2}K_0(\Delta \zeta)$, where, for this section, we will set
$\zeta = \sqrt{-s^2} = \sqrt{x^2 - (vt - i \alpha)^2}$. The limiting features of the Bessel function as its complex
argument is varied are \cite[Eqs.~\href{https://dlmf.nist.gov/10.31.E2}{(10.31.E), \href{https://dlmf.nist.gov/10.25.E3}{(10.25.E)}}]{DLMF}
\begin{align}
	\label{eq:asymp_1}
	K_0(\Delta \zeta) &\sim \ln\bigl(2/e^{\gamma}\Delta \zeta\bigr), &\quad\text{as $\zeta \to 0$},
\\
	\label{eq:asymp_2}
	K_0(\Delta \zeta) &\sim \sqrt{\frac{\pi}{2\Delta \zeta}} e^{-\Delta \zeta}, &\quad\text{as $|\zeta| \to \infty$},
\end{align}
where $\gamma = 0.577\dots$ is Euler's constant. Since, as before, we have to choose the principal branch of the
square root for $\zeta$ with $\Real \zeta >0$, we avoid the exponentially growing asymptotics of the Bessel function
at large, negative real arguments.

The basis of the following approximation is to seek a functional form that can interpolate between the asymptotics
of the Bessel function while keeping close to the correct shape, or the profile, throughout.
To facilitate this, we introduce the \emph{profile function} $P(\zeta)$ through the relation
\begin{equation}
	W_{0}(t,x) = q^{-1}\ln\bigl(P(\zeta)\bigr),
\end{equation}
where $q$ is a tuning constant to be chosen together with $P$. The task of approximating $W_{0}$ is then equivalent
to finding a suitable $P$ and $q$ pair. It is of course possible to precisely recover the original Bessel function
form through setting
\begin{equation}
	P(\zeta) = e^{\frac{q}{2}K_{0}(\Delta \zeta)},
\end{equation}
the asymptotics of which can be obtained through Eqs.~\eqref{eq:asymp_1} and \eqref{eq:asymp_2},
\begin{align}
	P(\zeta) &\sim \left(\frac{2}{e^{\gamma}\Delta\zeta}\right)^{q/2} &\quad\text{as $\zeta \to 0$},
	\label{eq:P_1}
\\[5pt]
	P(\zeta)
	&\sim e^{e^{-\Delta\zeta}\frac{q}{2}\sqrt{\frac{\pi}{2\Delta\zeta}}}
	&\quad\text{as $|\zeta| \to \infty$}.
	\label{eq:P_2}
\end{align}
Although these expressions are seemingly disconnected, we can take a series expansion of Eq.~\eqref{eq:P_2} and obtain
\begin{equation}
	 e^{e^{-\Delta\zeta}\frac{q}{2}\sqrt{\frac{\pi}{2\Delta\zeta}}}
	 = 1 + e^{-\Delta\zeta}\sqrt{\frac{\pi q^2}{8\Delta\zeta}} + \mathcal{O}\left(\frac{e^{-2\Delta\zeta}}{\Delta\zeta} \right).
\end{equation}
The algebraic decay in this expansion is $1/\sqrt{\zeta}$, whereas in Eq.~\eqref{eq:P_1} it is $1/\zeta^{q/2}$.
Therefore, with the choice of $q=1$ we can match the algebraic dependence in both limits. Furthermore $e^{-\Delta \zeta} \approx 1$
for small $\zeta$ such that $e^{-\Delta \zeta}/\sqrt{\zeta}$ produces an appropriate functional dependence in both limits,
albeit with inconsistent coefficients. It is clear then that a degree of tuning would be required to achieve a reasonable approximation,
and we aim for a compromise. Since in addition for small $\zeta$ we have $1/\sqrt{\zeta} \gg 1$, we propose the following profile function,
\begin{equation}
	P(\zeta)  = 1 + C \frac{e^{-\Delta\zeta}}{\sqrt{\Delta\zeta}},
	\label{eq:profile1}
\end{equation}
with $C$ a tuning parameter. Furthermore we shall keep a general $q$ as a tuning parameter since the
same $1/\sqrt{\zeta}$ dependence at large and small $\zeta$ comes to a price to a lack of accuracy
over the full range of $\zeta$. Thus, the profile function approximation we consider is
\begin{equation} \label{eq:W_prof}
	W_{0}(t,x) \approx W_{C,q}(t,x) = \frac{1}{q} \ln\left(1+\frac{Ce^{-\Delta\zeta}}{\sqrt{\Delta\zeta}} \right).
\end{equation}
\begin{figure}
	\centering
	\includegraphics[width=\columnwidth]{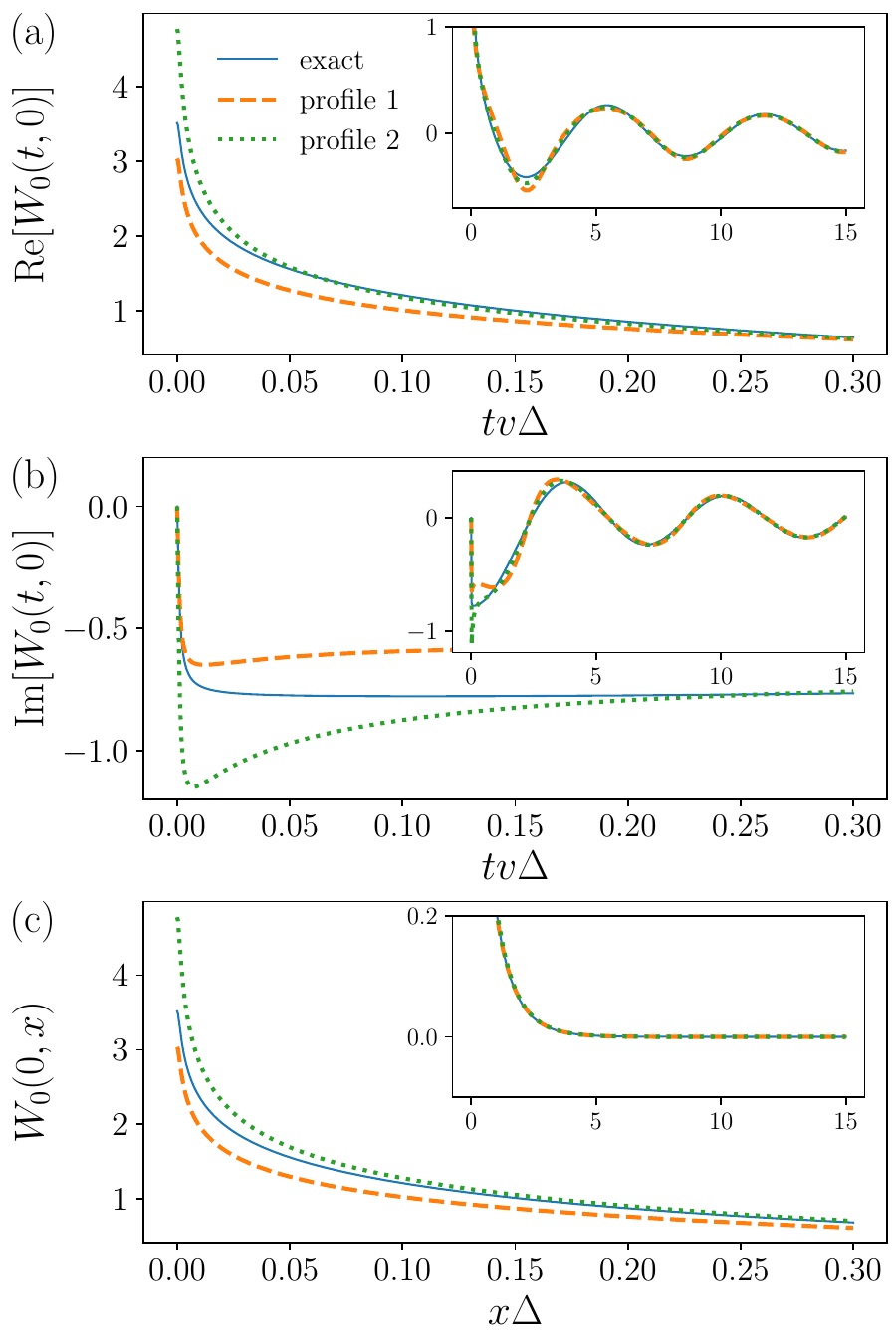}
	\caption{\label{fig:profile}%
	Comparison of the exact function $W_0(t,x)$ (solid lines) with two profile function approximations
	of Eq.~\eqref{eq:W_prof},
	``profile 1'' with $C = \sqrt{\pi/8}$ and $q=1$ (dashed lines),
	and ``profile 2'' with $C = \sqrt{\pi/32}$ and $q=1/2$ (dotted lines).
	Panel (a) shows the real and panel (b) the imaginary parts as a function of $t$ for $x=0$.
	Panel (c) shows the dependence on $x$ at $t=0$, for which all functions are real.
	The insets show the corresponding curves for a large range of $t$ and $x$ values, respectively.
	The cutoff is chosen as $\alpha = 10^{-3}/\Delta$.
	}
\end{figure}%
Figure~\ref{fig:profile} shows a comparison of the exact $W_0(t,x)$ with profile function approximations
for choices such that the curves at large arguments are perfectly matched. The seemingly obvious choice with
$q=1$ and consequently $C = \sqrt{\pi/8}$ (shown as dashed lines) has the right shape at $\zeta \to 0$ but due to the mismatch
of the amplitude $C$ fails to provide a decent description at small $\zeta$. For intermediate $\zeta$ values
the approximation follows the overall exact behavior but overestimates the curved structure.
By relaxing the $\zeta \to 0$ condition the overall aspect can be improved. This is shown by the dotted lines
in Fig.~\ref{fig:profile}, for the choice $q=1/2$ and $C = \sqrt{\pi q^2/8} = \sqrt{\pi/32}$. In this case
the profile function approximation converges with increasing $|\zeta|$ much faster to the exact curve, to the price of larger deviations
at small $\zeta$, which is particularly notable for $\Imag[W_0(t,x=0)]$ shown in panel (b).
This adjustment is still arbitrary but shows that the two parameters in the profile function approximation
provide some flexibility to tune the approximation to fit better some spacetime regions if required.
It seems also more natural to accept larger mismatches at short $\zeta$, since we have shown in Sec.~\ref{sec:GGE}
that there the appropriate description is given by a massless theory with the GGE imposed effective
temperature. The profile function cannot capture such a temperature behavior. Indeed one of the defining
features of a temperature dependent correlator is the periodicity in imaginary time with the inverse
Matsubara (or massubara for the GGE) frequencies. This replaces any $t$ dependence by $\sinh(\pi t/\beta)$
or $\sinh(\pi t/\beta_\text{eff})$ inside a $\log$ function that causes the latter to diverge.
Since for the profile function approximation we wish to maintain elementary functions only, we would either
be forced to replace the $\sinh$ by something more complicated, bringing us closer to the summations over
imaginary time shifted $K_0$ functions, or to replace the $\log$ by something more complicated.
Such changes would contradict
the idea behind the profile function to keep expressions as simple as possible. There is therefore not much
of a benefit to attempt extending this approximation to nonzero temperatures.

By adjusting $q$ and $C$ together it is possible to improve the profile function approximation for
certain spacetime regions, but setting $C = \sqrt{\pi q^2/8}$ always ensures that
the large $\zeta$ asymptotics is met exactly. The choice of $q=1/2$ and $C = \sqrt{\pi/32}$ made above
comes close to minimizing the mean square deviation between exact and profile functions over the
full interval $\zeta \in (0,\infty)$, which we find for $\alpha=10^{-3}/\Delta$
to be $q \approx 0.57$ with matching $C = \sqrt{\pi q^2/8}$.

To conclude this section, we emphasize that while the approximations provided in Secs.~\ref{sec:high} and \ref{sec:GGE}
for long and short spacetime separations were tightly bound to the physics of the fluctuations of the fields $\phi$,
the present discussion of the profile function follows more a convenient interpolation that captures the non-trivial
dependence on $\Delta$ and extends over all spacetime distances but has no deep physical motivation. As it captures
the qualitative features of the exact solution and has a simple logarithmic form, it may be of interest though if
having an analytic approximation is more useful than numerical accuracy.


\section{Discussion and Conclusions}
\label{sec:conclusions}

With the present paper we provide an extensive discussion of the basic correlation functions
of the 1+1 dimensional real, scalar, massive Klein-Gordon equation. Despite belonging to the oldest and most
elementary quantum field theories, a comprehensive discussion of the type provided has been
missing in the literature. We have started with a complete derivation of the exact
analytic form of the various correlation functions, with a focus on the Wightman function $W(t,x)$
as the basic building block. While such results can be found at numerous places in the literature
we provided here a complete derivation with special attention to the analytic properties arising
from the mass $\Delta$ dependence. We showed how an astute choice of a small imaginary shift
makes it possible to write down one single expression that is valid for all spacetime coordinates,
in the form of the modified Bessel function $K_0$,
with an analytic continuation between time-like and space-like regions. Fundamentally this requires
the imaginary shift to be infinitesimal but nonzero, but we gave arguments that practically the
shift should be promoted to the cutoff length $\alpha$ that regularizes the divergence on
the light cone. While this is not required for retarded and advanced Green's functions, and
not for the Wightman function off the light cone, it is a vital requirement for the
functions $G(t,x)$ that are shifted by $W(0,0)$ and arise as the natural correlators in the bosonization
of fermionic theories. While the temperature dependence had
to be included in terms of an infinite sum, we discussed various compact expressions
in the limits of high and low temperature, and of large and small spacetime arguments.

Most remarkable is that the correlators display for short spacetime distances transient effective
massless behaviors that are expressed in terms of generalized Gibbs ensembles, with effective
temperatures $T_\text{eff}(T)$ that interpolate between a purely mass-dependent quantity when
the physical temperature $T$ is small compared to the scale set by the mass and $T$ itself when
it is much larger. The generalized Gibbs ensemble is a modern development in the description of the steady state behavior of quenched
integrable systems that cannot thermalize due to their constants of motion. We showed that
this concept extends to the excitations on top of the ground state during the transient
regime, and that it is due to the maximal entanglement of the ground state between the
right and left moving modes that would not be correlated in the massless case.
Thus, the generalized Gibbs ensemble behavior is a characteristic of the qualitative
change of the ground state with and without a mass, and a clear manifestation
that a perturbative expansion in the mass can never provide an adequate description of the
correlators, however small the mass is. The distinction is also manifest with the
appearance of the mass dependent frequencies of the generalized Gibbs ensemble
that we named, not too seriously, the \emph{massubara} frequencies.
From the mathematical point of view the $T=0$ case also provides an approximation
to $K_0(z)$ at small $z$ that cannot be obtained from the regular series expansion.
With raising temperature we then showed that by adjusting the effective temperature
to include a real temperature dependence, the generalized Gibbs ensemble's validity
in the transient time regime extends to all temperatures and thus forms a robust feature
that compactly describes the effect of the increasing number of temperature correction
terms that need be added to the correlators.

In the final part of this paper we introduced a function that captures the main profile
of the exact solution by the tuning of two parameters. The purpose here was to replace
numerical accuracy by an elementary analytic approximation that, in the spirit of the
discussion above, captures the non-trivial dependence on $\Delta$ and provides a
decent approximation for all spacetime distances. Instead of being physically motivated
this approach was set up for the ease of further treatment of the correlation functions
if required.

In conclusion we have put forward a comprehensive derivation and analysis of the two-point
correlation functions for the Klein-Gordon theory in $1+1$ dimensions. This paper offers new insights
into this old problem with its exact solutions,
the finite temperature extension, the association of its features with fundamental
physical processes, and the approximation scheme.
With this paper we provide a reference for the fundamental two-point correlators that are the
building blocks for any physical application dominated by the fluctuations of a scalar field in 1+1 dimensions.
As alluded to in the introduction, such applications can be found in the broad spectrum from
the early universe to emergent many-body theories. Since the results are primarily useful
through this embedding in a broader context the accuracy of the expressions is central
to avoid the propagation of artifacts from approximations. Hence a large part of our discussion
is focused on the analytical properties of the correlators and their physical meaning, as well
as on the range of validity of the introduced approximations. We deem this as the most important
aspect of this work, and have given it full attention instead of some illustration on a single
example that may give some intuitive picture but misses the rigorous messages.

We anticipate that some of the results may be easily extended to higher dimensions,
as there is large precedence with exact formulas
\cite{Peskin1995,Mandl2010,Nair2005,Veltman1994,Zinn-Justin2002,Weinberg2005,Kaku1993,Huang2010,Roman1969,Itzykson1980,Greiner1992,diSessa1974,Zhang2010,Roepstorff1994,Mussardo2020}.
Furthermore it would be interesting to see how this physics is modified when models beyond
the Klein-Gordon theory are taken into account, such as the sine-Gordon model in which
alternative sophisticated methods exist to compute correlation functions \cite{Essler2005},
or within the context of the nonlinear Luttinger liquid \cite{Imambekov2012}.
For one-dimensional Bose gases the results could provide means to investigate further a conjectured connection
to the Kardar-Parisi-Zhang universality class \cite{Kulkarni2013}.
With our focus on the analytic properties of the correlation functions it would also
be interesting to explore an extension to general curved spacetime where there has been recent
progress in controlling analytic continuity in holomorphic formulations of field theories \cite{Guendelman2023}.


\begin{acknowledgements}

We thank P. Simon and G. Zarand for helpful discussions.
T. B. acknowledges the support from the EPSRC under Grant No.
EP/T518062/1. The work presented in this paper is theoretical. No data were produced,
and supporting research data are not required.

\end{acknowledgements}


\appendix


\section{Relations between the correlators}
\label{sec:identities}

The most straightforward way of obtaining a retarded correlator is to start by
computing the corresponding imaginary time (Matsubara) correlator. For a quadratic
action, the latter can be read off from the Euclidean version of the action's
kernel. In frequency space, where the imaginary time correlator depends on
the Matsubara frequencies $\omega_n$, one performs then the Wick rotation
$i\omega_n \to \omega + i 0$ to obtain the retarded correlator.
This procedure is straightforward and produces immediate results. It has,
however, two subtleties. First, it requires that there is an analytic continuation
in $\omega$ that allows for the Wick rotation. While mostly without problem, in
Ref. \cite{Braunecker2012} a situation was identified involving indeed scalar bosonic
fields in one spatial dimension in which an imaginary time Green's function was found to depend
on $|\omega_n|$, and thus encoded a nonanalytic behavior. We therefore perform a
direct computation, although at the end we can confirm that the Wick rotation remains
unproblematic in the present case. Second, the retarded function obtained by the
Wick rotation is temperature independent, and thus the spectral function is
temperature independent. This is expected but not evident from the general definition
and verifying this independence explicitly provides a good test of the validity of
the results. In the calculation below we repeat thus the standard textbook formal
manipulations of the Lehmann representations of the correlators, after which we
introduce the concrete evaluation of the correlators for the boson fields $\phi(\omega,k)$.

We consider two bosonic fields $B_1$ and $B_2$ and their retarded correlator
\begin{equation}
	G^r_{B_1,B_2}(t)
	= - i \hbar^{-1} \theta(t) \bmean{[B_1(t),B_2(0)]}.
\end{equation}
The evolution is governed by a Hamiltonian $H$, such that $B_{j}(t) = e^{i H t/\hbar} B_{j} e^{-i H t/\hbar}$ for $j=1,2$.
If the states $\ket{n}$ form a full set of eigenstates of $H$ with energies $E_n$, the thermal average is evaluated as
\begin{equation}
	\mean{\dots} = \frac{1}{Z}\sum_n e^{-\beta E_n} \bra{n} \dots \ket{n},
\end{equation}
with $Z = \sum_n e^{-\beta E_n}$.
For the retarded function this leads to
\begin{align}
	&G^r_{B_1,B_2}(t)
	= -i \hbar^{-1} \theta(t) \frac{1}{Z} \sum_{n,m} e^{-\beta E_n}
\nonumber\\
	&\times
	\biggl(
		e^{i (E_n-E_m)t/\hbar} \bra{n}B_1\ket{m} \bra{m}B_2\ket{n}
\nonumber\\
	&\qquad
	-
		e^{i (E_m-E_n)t/\hbar} \bra{n}B_2\ket{m} \bra{m}B_1\ket{n}
	\biggr)
\nonumber\\
	&= -i \hbar^{-1} \theta(t) \frac{1}{Z} \sum_{n,m}
	e^{i (E_n-E_m)t/\hbar}
	\bra{n}B_1\ket{m} \bra{m}B_2\ket{n}
\nonumber\\
	&\qquad
	\times
	\bigl(
		e^{-\beta E_n}
		-
		e^{-\beta E_m}
	\bigr).
\end{align}
Fourier transforming to frequency space yields
\begin{align}
	&G^r_{B_1,B_2}(\omega)
	= \frac{1}{Z} \sum_{n,m}
	\frac{\bra{n}B_1\ket{m} \bra{m}B_2\ket{n}}{\hbar\omega_+ + E_n-E_m}
\nonumber\\
	&\qquad
	\times
	\bigl(
		e^{-\beta E_n}
		-
		e^{-\beta E_m}
	\bigr),
\end{align}
with $\omega_+ = \omega + i\eta$, where $\eta>0$ represents an infinitesimal shift into the complex
plane to ensure convergence and causality of the result.

Using $\Imag[1/(x+i\eta)] = - \pi \delta(x)$, we identify the function $A_{B_1,B_2}(\omega)$
with the $\delta$ function dependent part
\begin{align}
	&A_{B_1,B_2}(\omega)
\nonumber\\
	&
	= \frac{\hbar}{Z} \sum_{n,m}
	\delta\bigl(\hbar\omega + E_n-E_m\bigr)
	\bra{n}B_1\ket{m} \bra{m}B_2\ket{n}
\nonumber\\
	&\qquad\quad
	\times
	\bigl(
		e^{-\beta E_n}
		-
		e^{-\beta E_m}
	\bigr)
\nonumber\\
	&
	=
	(1-e^{-\beta \hbar \omega})
	\frac{\hbar}{Z} \sum_{n,m}
	\delta\bigl(\hbar\omega + E_n-E_m\bigr)
\nonumber\\
	&\qquad\quad
	\times
	\bra{n}B_1\ket{m} \bra{m}B_2\ket{n}
	e^{-\beta E_n}.
\end{align}
If in addition $B_2 = B_1^\dagger$, as will be the case for all the results in this paper
we have
\begin{equation}
	A_{B_1,B_1^\dagger}(\omega) = -\frac{1}{\pi} \Imag G^r_{B_1,B_1^\dagger}(\omega),
\end{equation}
and $A_{B_1,B_1^\dagger}(\omega)$ is real, non-negative, and identified with the
spectral function.

If we consider now the correlator with fixed time order
\begin{equation}
	W_{B_1,B_2}(t) = \bmean{B_1(t) B_2(0)},
\end{equation}
we find in the same manner
\begin{align}
	&W_{B_1,B_2}(\omega)
	=
	2\pi \hbar \frac{1}{Z}
	\sum_{n,m}
	\delta\bigl(\hbar\omega + E_n-E_m\bigr)
\nonumber\\
	&\qquad\quad
	\times
	\bra{n}B_1\ket{m} \bra{m}B_2\ket{n}
	e^{-\beta E_n},
\end{align}
which provides the general identity
\begin{equation} \label{eq:fluct_diss_B}
	W_{B_1,B_2}(\omega) = \frac{2\pi\hbar}{1-e^{-\beta \hbar\omega}} A_{B_1,B_2}(\omega).
\end{equation}
To demonstrate the temperature independence of $G^r$ and $A$ we rely on two
conditions: that the action is Gaussian and that the operators $B_j$ ($j=1,2$)
will in \ref{sec:spectral} be identified with the boson fields $\phi$. The latter means
that $B_j$ can be decomposed in terms of the elementary creation and annihilation
operators $a_\ell^\dagger$ and $a_\ell$ of the harmonic modes as
\begin{equation}
	B_j = \sum_\ell \bigl( b_{j,\ell} a_{\ell} + c_{j,\ell} a_{\ell}^\dagger \bigr),
\end{equation}
with amplitudes $b_{j,\ell}$ and $c_{j,\ell}$. The summation symbol over $\ell$ is formal and can represent both a discrete
sum or an integral. From the Gaussian action it follows that
\begin{equation}
	B_j(t) = \sum_\ell \bigl( e^{-i \omega_\ell t} b_{j,\ell} a_{\ell} + e^{i \omega_\ell t} c_{j,\ell} a_{\ell}^\dagger \bigr),
\end{equation}
where $\hbar\omega_\ell$ are the elementary excitation energies of the modes $\ell$.
Using the canonical commutation relations $[a_\ell, a_{\ell'}^\dagger] = \delta_{\ell,\ell'}$ and
$[a_\ell, a_{\ell'}] = [a_\ell^\dagger,a_{\ell'}^\dagger]=0$, we have
\begin{equation} \label{eq:B1B2}
	[B_1(t),B_2(0)]
	= \sum_\ell \bigl( e^{-i \omega_\ell t} b_{1,\ell} c_{2,\ell} - e^{i \omega_\ell} c_{1,\ell} b_{2,\ell} \bigr),
\end{equation}
which is a number. Consequently
\begin{align} \label{eq:G_B1B2}
	G_{B_1,B_2}^r(t)
	&= -i  \hbar^{-1} \theta(t) [B_1(t),B_2(0)] \frac{1}{Z}\sum_n e^{-\beta E_n}
\nonumber\\
	&= -i  \hbar^{-1} \theta(t) [B_1(t),B_2(0)],
\end{align}
showing the temperature dependence indeed vanishes. By extension this demonstrates that $A_{B_1,B_2}(\omega)$ is
temperature independent too.


\section{Computation of the spectral function}
\label{sec:spectral}

For the translationally invariant system with a single field, the eigenmodes $\ell$ are
labeled by the momenta $k$ and the operators $a_k$. These are obtained by Legendre transforming
the action to the Hamiltonian
\begin{align}
	&H
	=
	\int \frac{dx}{2\pi}
\nonumber\\
	&\times
	\biggl\{
		\frac{v \pi^2}{\hbar} \Pi^2(x)
		+
		\hbar v
		\Bigl[
			\bigl(\partial_x \phi(x)\bigr)^2 +  \Delta^2 \phi^2(x)
		\Bigr]
	\biggr\},
\end{align}
with $\Pi = (\hbar/\pi v) \partial_t \phi$ the canonical momentum.
The change to Fourier modes is made through
\begin{align}
	\phi(x) &= \int_{-\infty}^\infty \frac{dk}{\sqrt{2\pi}} e^{i k x} \phi(k),
\\
	\Pi(x) &= \int_{-\infty}^\infty \frac{dk}{\sqrt{2\pi}} e^{i k x} \Pi(k),
\end{align}
and the eigenmodes diagonalizing the Hamiltonian are
\begin{align}
	\phi(k)
	&= \sqrt{\frac{\pi v}{2 \omega_k}}
	\bigl( a_k + a_{-k}^\dagger \bigr),
\label{eq:phi_a}
\\
	\Pi(k)
	&= -i\hbar\sqrt{\frac{\omega_k}{2\pi v}}
	\bigl( a_k - a_{-k}^\dagger \bigr),
\end{align}
with $\omega_k = v \sqrt{k^2 + \Delta^2}$ the eigenfrequencies,
such that $H = \int (dx/2\pi) \hbar \omega_k\,  a_k^\dagger a_k + \text{const}$.
Fourier transforming $G^r(t,x)$ of Eq.~\eqref{eq:Gr} with respect to $x$ leads to
\begin{align}
	&G^r(t,k) = -i\hbar^{-1} \theta(t) \int dk'
	\bmean{\bigl[\phi(t,k), \phi(0,k') \bigr]},
\end{align}
and to match Eq.~\eqref{eq:G_B1B2} we identify
$B_1 = \phi(t,k)$ and $B_2 = \phi^\dagger(0,k) = \phi(0,-k)$.
From Eq.~\eqref{eq:B1B2} with the generic $\ell$ replaced by $k$, and together with
Eq.~\eqref{eq:phi_a}
we obtain
\begin{equation}
	G^r(t,k) = -i\hbar^{-1} \theta(t) \frac{\pi v}{2 \omega_k} \bigl(
		e^{-i \omega_k t} - e^{i \omega_k t}
	\bigr),
\end{equation}
using $\omega_k = \omega_{-k}$. The further Fourier transform with respect to $t$
leads to
\begin{align}
	G^r(\omega,k) &= \frac{\pi v}{2 \hbar \omega_k} \left(
		\frac{1}{\omega_+-\omega_k} - \frac{1}{\omega_++\omega_k}
	\right)
\nonumber\\
	&= \frac{\pi v/\hbar}{\omega_+^2-\omega_k^2},
\end{align}
which provides the standard expression for the retarded correlator.
The spectral function follows then immediately from the identity $\Imag[1/(x+i\eta)] = -\pi \delta(x)$,
such that
\begin{align}
	A(\omega,k)
	&= -\frac{1}{\pi} \Imag G^r(\omega,k)
\nonumber\\
	&=
	\frac{\pi v}{2 \hbar \omega_k}
	\bigl[
		\delta(\omega-\omega_k) - \delta(\omega+\omega_k)
	\bigr].
\end{align}
With Eq.~\eqref{eq:fluct_diss}, or Eq.~\eqref{eq:fluct_diss_B},
this result allows for the computation of the further correlation
functions.


\section{Continuity of contour deformation}
\label{sec:contour_deformation_proof}

\begin{figure}
	\centering
	\includegraphics[width=0.7\columnwidth]{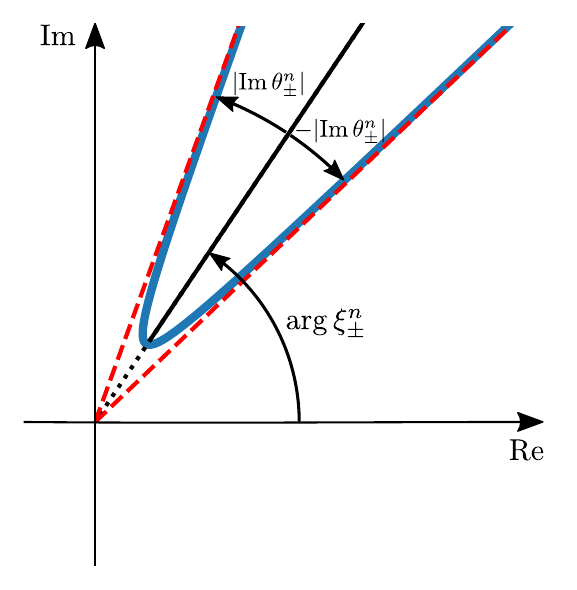}
	\caption{\label{fig:contour_temp_deform}%
		Illustration of the values of the exponent in the integral of Eq.~\eqref{eq:shifted_contour},
		represented in the complex plane by the thick solid blue curve. The curve is entirely confined to the
		positive half plane on a single Riemann sheet and extends between the asymptotes with angles
		$\arg\xi_\pm^n \pm \Imag\theta_\pm^n$ (dashed red lines). The central solid black line at angle
		$\arg\xi_\pm^n$ corresponds to the values of the exponent on the final contour in Eq.~\eqref{eq:shifted_contour_1}
		to which the initial curve can continuously be deformed.
	}
\end{figure}

The deformation of the integration contour from Eq.~\eqref{eq:shifted_contour} to Eq.~\eqref{eq:shifted_contour_1}
requires that the exponential $e^{-\xi_\pm^n \cosh(\varphi)}$ in the integrand remains convergent, such
that $\Real\left[\xi_\pm^n \cosh(\varphi)\right] > 0$, throughout the deformation.
As the integration runs over $\Real\varphi$ we need thus to verify if $\Imag\varphi$ can be smoothly connected
from its initial value $\Imag\varphi = \Imag\theta_\pm^n$ to its final value $\Imag\varphi = 0$.

As $\xi_\pm^n = \Delta \sqrt{x^2 - v^2(t\pm i n \beta \hbar)^2}$ it is always possible to choose the square root
such that its real part is positive. Furthermore we can choose the branch of the square root such that
$|\arg(\xi_\pm^n)| < \pi/2$, and that consequently the final contour integration lies entirely on the
positive real half plane of the first Riemann sheet. A continuous deformation of the integration contour
is then possible if the initial contour lies entirely in the same half plane of the same Riemann sheet.

For the initial contour we write $\varphi = \varphi' + i \Imag\theta_\pm^n$ for real $\varphi'$.
At $\varphi'=0$ we have $\arg[\cosh(i \Imag\theta_\pm^n)] = 0$ if
we choose the principal branch of the $\tanh^{-1}$ that defines $\theta_\pm^n$. Consequently
$\arg[ \xi_\pm^n \cosh(\varphi)]$ at $\varphi'= 0$ is identical to the $\arg(\xi_\pm^n)$ of the final contour,
such that continuity at this point is already guaranteed. Since by construction of $\xi_\pm^n$ and $\theta_\pm^n$
from Eq.~\eqref{eq:W_n_2} the exponential in the integrand is convergent for all $\varphi'$ the argument
of $\xi_\pm^n \cosh(\varphi)$ cannot pass through $\pm \pi/2$ such that the proof is completed.

To provide some more detail, we note that $\arg[\cosh(\varphi'+i \Imag\theta_\pm^n)]$
varies smoothly and monotonously between the asymptotes $\pm \Imag\theta_\pm^n$ for $\phi' \to \pm \infty$,
which are determined by the behavior $\cosh(\phi' + i \Imag\theta_\pm^n) \sim e^{\pm (\phi' + i \Imag\theta_\pm^n)}/2$
in these limits. Hence, $\arg[\xi_\pm^n \cosh(\phi'+i \Imag\theta_\pm^n)$ lies within the
interval spanned by $\arg(\xi_\pm^n) \pm \Imag\theta_\pm^n$, as shown in Fig. \ \ref{fig:contour_temp_deform}.
Again by construction of $\xi_\pm^n$ and $\theta_\pm^n$
the condition $|\arg(\xi_\pm^n) \pm \Imag\theta_\pm^n| < \pi/2$ is satisfied,
so that all values of $\xi_\pm^n \cosh(\varphi)$ are confined
entirely to the positive real half plane of the first Riemann sheet.
As a consequence there exists a continuous deformation of the integration contour between the intended
initial and final shapes that maintains strict convergence throughout, and consists in collapsing the
curve between two asymptotes indicated by the dashed lines in Fig.~\ref{fig:contour_temp_deform} onto the
central black line.


\section{Details for generalized Gibbs ensemble for gapped one-dimensional system}
\label{sec:massive_Dirac}

In this appendix we complete the general discussion of Sec.~\ref{sec:GGE} with
explicit calculations for the massive Dirac model equivalent to the considered
sine-Gordon model. The reason for splitting this off to this appendix is that
the calculations to follow are too simple to allow gaining general insight, yet
they are necessary to confirm the general assumptions made in the main text.

Let for a fixed momentum $k$ the Dirac theory be described by right $(+)$ and left $(-)$
moving fermions, denoted by the creation and annihilation operators $c_\pm^\dagger, c_\pm$,
and subject to the
Hamiltonian
\begin{equation}
	H_k = \hbar v k \bigl( c_+^\dagger c_+ - c_-^\dagger c_- \bigr)
	+ \Delta \bigl( c_+^\dagger c_- + c_-^\dagger c_+ \bigr).
\end{equation}
To be able to perform partial traces we need to work in the full $4 \times 4$ Hilbert
space of states $\ket{n_+,n_-}$ with the occupation numbers $n_\pm = 0,1$ such that
$c_\pm^\dagger c_\pm \ket{n_+,n_-} = n_\pm \ket{n_+,n_-}$.
We treat
$H_\Delta =	\Delta \bigl( c_+^\dagger c_- + c_-^\dagger c_+ \bigr)$
as the unperturbed Hamiltonian setting the large energy scale. Its
ground state with energy $E_0 = -\hbar v \Delta$
is obtained for the one-fermion wave function $\ket{\Phi_0} = \frac{1}{\sqrt{2}}(\ket{1,0} - \ket{0,1})$
and, as the full Hamiltonian is particle number conserving, there is only one
excited state reachable, $\ket{\Phi_1} = \frac{1}{\sqrt{2}}(\ket{1,0} + \ket{0,1})$,
with energy $E_1 = +\hbar v \Delta$.

The perturbation is given by $H_\pm  = \pm \hbar v k c_\pm^\dagger c_\pm$
for which we observe that the matrix elements between $\ket{\Phi_0}$ and $\ket{\Phi_1}$
are identical,
\begin{equation}
	\bra{\Phi_1} H_+ \ket{\Phi_0} = \frac{1}{2} \bra{1,0} (\hbar v k) \ket{1,0} = \frac{\hbar v k}{2},
\end{equation}
and
\begin{equation}
	\bra{\Phi_1} H_- \ket{\Phi_0} = \frac{1}{2} \bra{0,1} (-\hbar v k) (-\ket{0,1}) = \frac{\hbar v k}{2}.
\end{equation}
This confirms that Eq.~\eqref{eq:Psi_0} holds.
From the full initial density matrix $\rho^0 = \ket{\Psi_0} \bra{\Psi_0} = \frac{1}{2} (\ket{1,0} - \ket{0,1})(\bra{1,0} - \bra{0,1})$
we obtain the reduced density matrices
\begin{equation}
	\rho^0_\pm = \Tr_\mp\{\rho^0\}
	= \frac{1}{2} \bigl( \ket{0}\bra{0}_\pm + \ket{1}\bra{1}_\pm ) = \frac{1}{2} \mathbbm{1}_\pm,
\end{equation}
which are each proportional
to the identity in their corresponding $\pm$ subspace.
This corroborates the condition of maximal entanglement required in Sec.~\ref{sec:GGE}.
Since each state can be occupied or empty, $N_\pm = 2$.


%


\end{document}